\documentclass[12pt, a4paper]{article}
\pdfoutput=1
\usepackage{graphicx}
\usepackage{tikz}
\usetikzlibrary{decorations.pathreplacing,decorations.markings,snakes}
\usetikzlibrary{decorations.pathmorphing}
\tikzset{snake it/.style={decorate, decoration=snake}}
\usepackage{jheppub}
\usepackage{amsmath,amssymb,euscript,array,mathrsfs,appendix,ctable,mathabx}
\usepackage{bm}
\usepackage[small]{caption}
\usepackage[dvipsnames]{xcolor}
\usepackage{float}
\usepackage{braket}
\usepackage{comment}
\usepackage{xcolor}
\usepackage{empheq}
 \usepackage{slashed} 
\definecolor{lightblue}{RGB}{100,180,255}
\newcommand \arXiv [1]{\href{http://arxiv.org/abs/#1}{\tt arXiv:#1}}

\def\ben
{\begin{equation}}
\def\een{\end{equation}}

    \let\L=\Lambda
   
 \let\W=\mu

\def\W={\cal W}
\def\L ={\cal L}

\def\be{\begin{equation}}
\def\ee{\end{equation}}
\def\ba{\begin{array}}
\def\ea{\end{array}}

\def\dalemb#1#2{{\vbox{\hrule height .#2pt
        \hbox{\vrule width.#2pt height#1pt \kern#1pt
                \vrule width.#2pt}
        \hrule height.#2pt}}}

\newcommand{\bea}{\begin{eqnarray}}
\newcommand{\eea}{\end{eqnarray}}

\title{Entanglement Revivals  and Scrambling for Evaporating Black Holes}
\author{Levy B. N. Batista$^a$, Nicol\`o Bragagnolo$^a$, Rhys Holmes$^b$ and S. Prem Kumar$^a$}
\affiliation{$^{(a)}$Centre for Quantum Fields and Gravity,\\ Department of Physics, Swansea University, \\Singleton Park, Swansea, SA2 8PP, U.K.}
\affiliation{$^{(b)}$ School of Physics and Astronomy, Cardiff University, \\5 The Parade, Cardiff, CF24 3AA, U.K.}
\emailAdd{L.DNBatista@swansea.ac.uk,nicolo.bragagnolo@swansea.ac.uk, HolmesRA1@cardiff.ac.uk, s.p.kumar@swansea.ac.uk}

\abstract{We investigate the spreading of entanglement, and entanglement memory effects, in two-dimensional conformal field theory (CFT) propagating on evaporating black hole backgrounds. Memory effects leading to late-time spikes in mutual information for widely separated intervals are well known in CFTs admitting a quasiparticle description. In this work we examine the effect of black hole scrambling on late time mutual information spikes for disjoint intervals in free fermion CFT prepared in a thermofield double state.
Late-time entanglement revival is driven by island-induced purification of modes in the union of the intervals. We show across two distinct 2d gravity models, Jackiw-Teitelboim (JT) gravity and the Russo-Susskind-Thorlacius (RST) model, that parametrically dialing up black hole scrambling time smooths out and suppresses entanglement spikes until they disappear at a critical scale, interpolating between free quasiparticle and maximal scrambling pictures. At the critical point, the interval lengths are exponential in black hole scrambling time. We further find a very closely related effect manifest as an entanglement dip for a single interval in a single-sided evaporating RST black hole.
}

\begin{document}
\maketitle

\section{Introduction}
The scrambling of entanglement is a hallmark of quantum chaotic systems. First enunciated in the context of information loss and retrieval from black holes \cite{Hayden:2007cs,Sekino:2008he, Shenker:2013pqa}, how entanglement spreads and scrambles in quantum field theories (QFTs) is an interesting question in its own right. For QFTs in which excitations have a quasiparticle description, entanglement built into an atypical initial state for the quasiparticles can survive at relatively long separations and at late times \cite{Asplund:2015eha}. A particularly useful and striking diagnostic test of this is provided by the time evolution of the mutual information between two well separated finite intervals in 1+1-dimensional CFT \cite{Asplund:2015eha, Calabrese:2009qy} which exhibits a late time spike. Such spikes are however absent in large-$c$ holographic CFTs which are maximally scrambling \cite{Asplund:2013zba,Leichenauer:2015xra}.

Evaporating black holes bring a new element into the mix, as the black hole will additionally scramble the correlations built into the initial QFT matter state. Effective 1+1-dimensional models such as Jackiw-Teitelboim (JT) gravity \cite{Jackiw:1984je, Teitelboim:1983ux} coupled to a nongravitating CFT bath and the Russo-Susskind-Thorlacius (RST) model \cite{Russo:1992ax, Fiola:1994ir}  provide a calculable framework to examine entanglement scrambling in general, by employing the generalised entropy and island formalism \cite{Almheiri:2019yqk, Almheiri:2019qdq, Penington:2019kki,Hartman:2020swn}. In this situation, there are potentially two time scales at play -- the black hole scrambling time (at temperature $\beta^{-1}$),\footnote{The extra factor of $c$ in the argument of the log arises from the fact that there are $c$ CFT degrees of freedom and therefore $t_{\rm scr}$ is correspondingly shorter \cite{Almheiri:2019yqk}. }
\be
t_{\rm scr}=\frac{\beta}{2\pi}\log \frac{\Delta S_{\rm BH}}{c}\,,
\ee
where $\Delta S_{\rm BH}$ is the entropy excess above the extremal value \cite{Leichenauer:2014nxa},
and if the CFT  is chaotic with a Lyapunov exponent $\lambda_L$ then a CFT scrambling time scale $t_{\rm CFT}=\frac{1}{\lambda_L}\log c$. Free CFTs, and more generally CFTs with infinite higher spin conserved currents have vanishing Lyapunov exponents and correspondingly do not scramble ($t_{\rm CFT}\to\infty$) \cite{Perlmutter:2016pkf}.

In this paper, we study free fermion CFT propagating in a 1+1-dimensional black hole background. Our motivation stems from the scenario originally examined in \cite{Hollowood:2021wkw} for the two-sided AdS$_2$ black hole in JT gravity coupled to Minkowski CFT baths in the thermofield double (TFD) state. In that work, the mutual information evolution of two disjoint finite intervals $A_L$ and $A_R$, placed in the left and right copies of the CFT baths showed a peak for well separated intervals at late times.\footnote{For simplicity, we consider only the symmetric case, i.e., when the intervals $A_L$ and $A_R$ are identical.} In particular, at high enough temperatures and for large enough intervals, so $t_{\rm scr}$ is negligible in the limit, the entanglement entropy $S(A_L \cup A_R)$  behaves identically to that of free fermion CFTs on the half-line, ${\rm BCFT}_L$ and ${\rm BCFT}_R$, prepared in a TFD state with the boundaries acting as reflecting mirrors for the quasiparticles. $S(A_L \cup A_R)$ initially grows until it saturates at the corresponding thermal result. The inclusion of an island accounts for an entropy dip (equivalently, a spike in the mutual information $I(A_L, A_R)$) {\em well after} entanglement equilibrium has been reached. This purification can be quantitatively captured via a geometric or ray optics description of Hawking modes entering the region $A_L\cup A_R$ at the same time as their maximally entangled partner modes enter the island, or equivalently in the BCFT picture,  when the partner modes enter $A_L \cup A_R$ after reflection from the boundary mirror.

We will examine below how the picture above is affected by the finite scrambling time of the black hole. Across the two setups - the AdS$_2$ black hole within JT gravity, and the asymptotically flat black hole in the RST model -  we find that the mutual information spike is smoothed out and reduced by the black hole scrambling effect, disappearing entirely below a critical value of the interval size,
\begin{equation}\label{crituniversal}
    \boxed{L_{\rm crit}=c_{1}\frac{\beta}{2\pi}e^{\frac{2\pi}{\beta}t_{\text{scr}}}\;+ c_{2}t_{\text{scr}}\,+\mathcal{O}(\beta)\,,}
\end{equation}
where $c_{1,2}$ are $\mathcal{O}(1)$ dimensionless theory- and state-dependent coefficients, and we work to leading order in a high temperature expansion $\beta \ll t_{\rm scr}, L$ where $L$ is the interval size. As a bonus we find an additional critical scale for large intervals wherein a transient period prior to the entanglement the entropy dip (dubbed regime (II) in \cite{Hollowood:2021wkw}) disappears as the scrambling time is dialled up.

We further find a closely related phenomenon - an entanglement entropy dip in a single, finite interval of radiation in the single-sided evaporating black hole in the RST model. We find that the dip in this case is also smoothed out and disappears below a critical interval length given by \eqref{crituniversal}.

Our paper is organised as follows: In section \ref{sec2} we review the well studied setup of JT gravity coupled to nongravitating CFT baths, and proceed to the analytical evaluation of the mutual information of two intervals in the TFD state and plotting the results for different numerical values of $\beta$, scrambling time,  and interval length. Section \ref{sec3} deals with the RST model in entirety. We first review general aspects of the RST model, and discuss the phenomenon of entanglement dips in the single-sided evaporating background, moving on finally to the eternal black hole solution to study entanglement revivals in the TFD state. We conclude with a summary and directions for future work in section \ref{sec4}.

\section{JT black hole coupled to Minkowski radiation baths}
\label{sec2}
We first examine the eternal black hole in AdS$_2$ within JT gravity \cite{Jackiw:1984je, Teitelboim:1983ux}, coupled to two non-gravitating Minkowski reservoirs. A 2d dilaton gravity model, JT gravity describes the low-energy dynamics near the horizon of near-extremal black holes in four dimensions. The 2d dilaton is proportional to the area of the two-sphere in the higher  dimensional setting \cite{Mertens:2022irh}, and its value at the horizon determines the  Bekenstein-Hawking (BH) entropy of the black hole.

In this setup, the black hole is in thermal equilibrium with the radiation bath whose degrees of freedom are described by a CFT, which we take to be the free fermion theory for definiteness. These massless fields propagate across the AdS and Minkowski regions with transparent boundary conditions on the regularised AdS boundary that is glued onto the non-gravitating region.
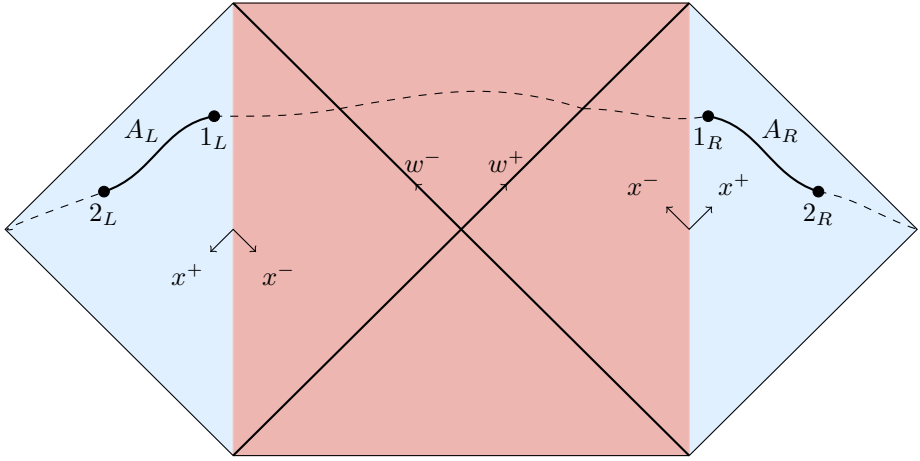
\begin{figure}[ht]
\begin{center}
\begin{tikzpicture}[scale=1]
\draw[fill=lightblue!20,lightblue!20] (-6,0) -- (-3,-3) -- (-3,3) -- cycle;
\draw[fill=lightblue!20,lightblue!20] (6,0) -- (3,-3) -- (3,3) -- cycle;
\draw[pink,fill=Plum!10!pink] (-3,-3) -- (3,-3) -- (3,3) -- (-3,3) -- cycle;
\draw[-] (-3,3) -- (-6,0) -- (-3,-3);
\draw[-]  (3,-3) -- (6,0) -- (3,3);
\draw[-] (-3,-3) -- (3,-3);
\draw[-] (-3,3) -- (3,3);
\draw[thick] (-3,-3) -- (3,3);
\draw[thick] (-3,3) -- (3,-3);
\filldraw[black] (3.25,1.5) circle (2pt);
\filldraw[black] (-3.25,1.5) circle (2pt);
\filldraw[black] (4.7,0.5) circle (2pt);
\filldraw[black] (-4.7,0.5) circle (2pt);
\node at (-4.7,0.2) {\footnotesize $2_L$};
\node at (-3.25,1.2) {\footnotesize $1_L$};
\node at (3.25,1.2) {\footnotesize $1_R$};
\node at (4.7,0.2) {\footnotesize $2_R$};
\draw[dashed] (-5.9,0) to[out=-170,in=18] (-4.7,0.5);
\draw[thick] (-3.25,1.5) to[out=-170,in=18] (-4.7,0.5);
\draw[dashed] (-3.25,1.5) to[out=0,in=190] (-1.6,1.6);
\draw[dashed] (-1.6, 1.6) to[out=10,in=162] (1.6, 1.6);
\draw[dashed] (1.6, 1.6) to[out=0,in=190] (3.25,1.5);
\draw[thick] (3.25,1.5) to[out=-10,in=162] (4.7,0.5);
\draw[dashed] (4.7,0.5) to[out=-10,in=162] (6,0);
\node at (4.2,1.3) {\footnotesize $A_R$};
\node at (-4.2,1.3) {\footnotesize $A_L$};
%
%\node at (3.3,3.3) {\footnotesize $w^+$};
%\node at (-3.3,3.3) {\footnotesize $w^-$};
%
\draw[<->] (2.7,0.3) -- (3,0) -- (3.3,0.3);
\node at (2.4,0.6) {\footnotesize $x^-$};
\node at (3.6,0.6) {\footnotesize $x^+$};
\draw[<->] (-0.6,0.6) -- (0,0) -- (0.6,0.6);
\node at (-0.5,0.9) {\footnotesize $w^-$};
\node at (0.6,0.9) {\footnotesize $w^+$};
\begin{scope}[yshift=0cm]
\draw[<->] (-2.7,-0.3) -- (-3,0) -- (-3.3,-0.3);
\node at (-2.4,-0.6) {\footnotesize $x^-$};
\node at (-3.6,-0.6) {\footnotesize $x^+$};
\end{scope}
%
%\draw[<->] (0.3,1.2) -- (0,1.5) -- (0.3,1.8);
%\node at (0.6,0.9) {\footnotesize $x^-$};
%\node at (0.6,2.1) {\footnotesize $x^+$};
%
\end{tikzpicture}
\caption{\footnotesize The Penrose diagram describing an eternal black hole in AdS$_2$ glued onto two half Minkowski baths. The intervals $A_L$ and $A_R$ are contained in the non-gravitating regions and are part of the same Cauchy slice (dashed) in the resulting spacetime.}
\label{fig1} 
\end{center}
\end{figure}

\subsection{Setup and notation}
Let us now set the notation for the rest of this section. As shown in figure \ref{fig1} above we consider two intervals, $A_L$ with endpoints  $2_L$ and $1_L$, and $A_R$ whose endpoints are $1_R$ and $2_R$, defined on the same Cauchy slice. We take Minkowski bath coordinates $(t, x)$ with spatial coordinate $x \geq 0$ and $t$, the bath time. They can be extended to the AdS region outside the black hole by taking $x \in \mathbb{R}$, and identifying $t$ with Schwarzschild time. We then define appropriate null coordinates,
\begin{equation}
    x^\pm_R = t \pm x\,, \qquad x^\pm_L = -t \pm x + \frac{i\beta}{2}\,,
\end{equation}
on the right $(R)$ and left $(L)$ regions respectively. As required for the Hartle-Hawking initial state, we have taken $t_L = - t_R+ i\beta/2$. These lead to  Kruskal-Szekeres (KS) coordinates that cover the whole spacetime,
\begin{equation}\label{Kruskal}
    w^\pm = \pm e^{\pm 2\pi x^\pm/\beta}.
\end{equation}
The left and right (future) horizons are at $w^+ = 0$ and $w^- = 0$, respectively, and the  AdS$_2$ metric in these coordinates is,
\begin{equation}
    ds^2 = -\frac{4dw^+dw^-}{(1 + w^+w^-)^2}\,.
\end{equation}
The AdS$_2$ conformal boundaries are at $w^+w^- = -1$, and the dilaton profile is given by
\begin{equation}
    \phi = \phi_0 + \frac{2\pi \phi_r}{\beta}\frac{1 - w^+w^-}{1+w^+w^-}.
\end{equation}
In JT gravity, the BH entropy of the black hole is given by the value of the dilaton at the horizon,
\begin{equation}
    S_{\text{BH}} = \frac{\phi}{4G_N}\bigg|_{w^- = 0} = S_0+\frac{\pi c }{6\beta k},
\end{equation}
where $k \equiv G_Nc/3\phi_r$ is the JT parameter, and $S_0=\frac{\phi_0}{4G_N}$ is the extremal value of the entropy. The parameter $k$ has dimensions of inverse length and in units where the AdS radius is set to one, $k\ll 1$
to ensure the validity of the semiclassical approximation.
\subsection{Entanglement revivals and critical disappearance}

To understand how the information spreads across spacetime, we consider the mutual information between the intervals $A_L$ and $A_R$,
\begin{equation}
    I(A_L, A_R) = S(A_L) + S(A_R) - S(A_L\cup A_R),
\end{equation}
where $S(A)$ is the entanglement entropy associated to an interval $A$. The mutual information provides a measure of how correlated $A_L$ and $A_R$ are. Furthermore, unlike entanglement entropy it is free of UV divergences.

To compute this mutual information in the system coupled to dynamical gravity we need the individual fine-grained or generalised entropies for $A_{L,R}$ and $A_L\cup A_R$. For any interval $A$, this is given by the island formula,
\begin{equation}\label{islandsDef}
    S(A) = \text{min}\,\text{ext}_{\partial I}\Big[\sum_{\partial I}\frac{\text{Area}(\partial I)}{4G_N} + S_{\text{CFT}}(I \cup A)\Big],
\end{equation}
where we admit the inclusion of additional intervals $I$, the islands, with boundaries $\partial I$, so-called quantum extremal surfaces (QES) corresponding to points in the 2d effective gravity models.  $S_{\text{CFT}}(I\cup A)$ accounts for the semiclassical von Neumann entropy of the quantum fields on the region $I \cup A$.\footnote{In principle, nothing prevents the extremisation in eq. \eqref{islandsDef} from resulting in multiple island saddles. Thus, we complete the prescription by picking the one that minimises the generalised entropy. } The CFT computation of this is standard, and for free fermions the exact result for arbitrary choices of intervals is known (see \cite{Casini:2009sr} for a review).

\subsubsection{No-island saddle and late times} 
Given the symmetric choice of the intervals $A_{L,R}$, the KS coordinates of their endpoints $2_L$, $1_L$, $1_R$, and $2_R$ are,
\begin{equation}
    \begin{cases}
        w^{\pm}_{1R} = w^{\mp}_{1L} = \pm e^{2\pi(\pm t + a)/\beta}, \\
        w^{\pm}_{2R} = w^{\mp}_{2L} = \pm e^{2\pi(\pm t + b)/\beta}.
    \end{cases}
\end{equation}
By construction, we have that $S(A_L) = S(A_R)$. In the TFD state all single-sided correlators are given by their thermal equilibrium values. Entanglement and its growth only manifests in two-sided observables and correlation functions. Further, inclusion of a potential island contribution only increases the fine-grained entropy, as the area term represents a large contribution. Therefore,
\begin{equation}\label{EntropySingleInterval}
    S(A_R) = S(A_L)= \frac{c}{3}\log \sinh \tfrac{\pi}{\beta} L\,,\qquad L\equiv b-a\,.
\end{equation}
For $S(A_L \cup A_R)$ on the other hand, we must consider the competition between the island and no-island (Hawking) saddles. In the  Hawking or no-island saddle for late times after a short period of linear growth $S_\emptyset(A_L\cup A_R)$ approaches the equilibrium value\footnote{This is valid when small intervals with $b<3a$, while for $b>3a$ the island saddle will possibly affect also the early-growth regime. This will be discussed in detail in \ref{b>3aJT}.} determined by eq.\eqref{EntropySingleInterval} (see \cite{Hollowood:2021wkw} for exact expressions),
\be
S_\emptyset(A_L\cup A_R) \to 2 S(A_{L,R})\,.
\ee
Therefore the mutual information vanishes at late times in the no-island saddle.
On the other hand the island contribution to $S(A_L\cup A_R)$  results in a late-time dip in the entropy and spike in the mutual information for times $a< t< b$ as demonstrated in \cite{Hollowood:2021wkw} in the high temperature limit $L/\beta\gg 1$, and when the scrambling time is also negligible $a, b, L\gg t_{\rm scr}$. 

\subsubsection{Black hole scrambling time}
In the present analysis our goal is to understand the effect of finite scrambling time on the entanglement dip or mutual information spike. To this end, it will be useful and natural to define the small dimensionless parameter $s$,
\be
s\equiv\frac{\beta k}{2\pi }\ll1\,,
\ee
in terms of which the black hole scrambling time can be expressed as,
\begin{equation}\label{SystemRegime}
     t_{\text{scr}} = \frac{\beta}{2\pi} \log \frac{\Delta S_{\rm BH}}{c}=\frac{\beta}{2\pi} \left[\log \,\frac{1}{s} + {\cal O}(1)\right]\,.
\end{equation}
A quick route to this result for the scrambling time is by calculating how much boundary time has elapsed before a null diary sent in from the AdS$_2$ boundary reaches the QES and becomes recoverable in the radiation. 

We now want to work in a regime where scrambling time is non-negligible compared to other length and time scales $L, a,b, t$, and is parametrically larger than the thermal time 
\be
 t_{\rm scr}\gg \frac{\beta}{2\pi}\,,\label{highT}
\ee
%where $d \sim L, a, b, t$. 
Note that small $s$ automatically implies that scrambling time is much bigger than the thermal time scale. Below, we explain what happens to the mutual information $I(A_L, A_R)$ as we vary $s$, $L$, and $\beta$.
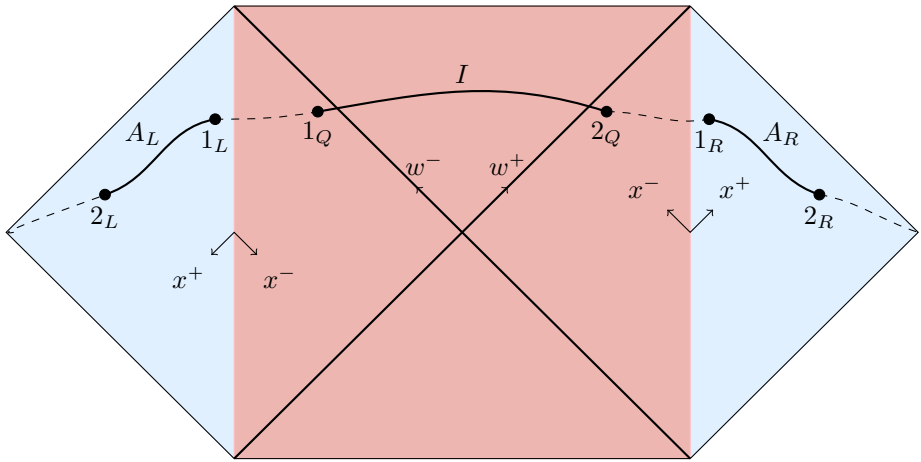
\begin{figure}[ht]
\begin{center}
\begin{tikzpicture}[scale=1]
\draw[fill=lightblue!20,lightblue!20] (-6,0) -- (-3,-3) -- (-3,3) -- cycle;
\draw[fill=lightblue!20,lightblue!20] (6,0) -- (3,-3) -- (3,3) -- cycle;
\draw[pink,fill=Plum!10!pink] (-3,-3) -- (3,-3) -- (3,3) -- (-3,3) -- cycle;
\draw[-] (-3,3) -- (-6,0) -- (-3,-3);
\draw[-]  (3,-3) -- (6,0) -- (3,3);
\draw[-] (-3,-3) -- (3,-3);
\draw[-] (-3,3) -- (3,3);
\draw[thick] (-3,-3) -- (3,3);
\draw[thick] (-3,3) -- (3,-3);
\filldraw[black] (1.9,1.6) circle (2pt);
\filldraw[black] (3.25,1.5) circle (2pt);
\filldraw[black] (-3.25,1.5) circle (2pt);
\filldraw[black] (4.7,0.5) circle (2pt);
\filldraw[black] (-4.7,0.5) circle (2pt);
\filldraw[black] (-1.9,1.6) circle (2pt);
\node at (-4.7,0.2) {\footnotesize $2_L$};
\node at (-3.25,1.2) {\footnotesize $1_L$};
\node at (3.25,1.2) {\footnotesize $1_R$};
\node at (4.7,0.2) {\footnotesize $2_R$};
\node at (1.9, 1.3) {\footnotesize $2_Q$};
\node at (-1.9, 1.3) {\footnotesize $1_Q$};
\draw[dashed] (-5.9,0) to[out=-170,in=18] (-4.7,0.5);
\draw[thick] (-3.25,1.5) to[out=-170,in=18] (-4.7,0.5);
\draw[dashed] (-3.25,1.5) to[out=0,in=190] (-1.9,1.6);
\draw[thick] (-1.9, 1.6) to[out=10,in=162] (1.9, 1.6);
\draw[dashed] (1.9, 1.6) to[out=0,in=190] (3.25,1.5);
\draw[thick] (3.25,1.5) to[out=-10,in=162] (4.7,0.5);
\draw[dashed] (4.7,0.5) to[out=-10,in=162] (6,0);
\node at (4.2,1.3) {\footnotesize $A_R$};
\node at (-4.2,1.3) {\footnotesize $A_L$};
\node at (0, 2.1) {\footnotesize $I$};
%
%\node at (3.3,3.3) {\footnotesize $w^+$};
%\node at (-3.3,3.3) {\footnotesize $w^-$};
%
\draw[<->] (2.7,0.3) -- (3,0) -- (3.3,0.3);
\node at (2.4,0.6) {\footnotesize $x^-$};
\node at (3.6,0.6) {\footnotesize $x^+$};
\draw[<->] (-0.6,0.6) -- (0,0) -- (0.6,0.6);
\node at (-0.5,0.9) {\footnotesize $w^-$};
\node at (0.6,0.9) {\footnotesize $w^+$};
\begin{scope}[yshift=0cm]
\draw[<->] (-2.7,-0.3) -- (-3,0) -- (-3.3,-0.3);
\node at (-2.4,-0.6) {\footnotesize $x^-$};
\node at (-3.6,-0.6) {\footnotesize $x^+$};
\end{scope}
%
%\draw[<->] (0.3,1.2) -- (0,1.5) -- (0.3,1.8);
%\node at (0.6,0.9) {\footnotesize $x^-$};
%\node at (0.6,2.1) {\footnotesize $x^+$};
%
\end{tikzpicture}
\caption{\footnotesize In the generalised entropy proposal, one of the possible saddles contains an island $I$, which will ``perturb'' the entanglement equilibrium of region $A_L \cup A_R$ at late times, exhibiting a memory effect when the black hole scrambling is subdominant.}
\label{fig2} 
\end{center}
\end{figure}

\subsubsection{The critical length}\label{critical_length_analysis}
First we will show that there is a critical size for the length of the intervals $A_{L,R}$ below which the dip in the entanglement entropy of $A_L \cup A_R$ will disappear. The idea is to find the minimum entropy as described by the island saddle and see under which conditions this is larger than the thermal entropy of both intervals. The minimal island saddle has two QESs, which we denote by $1_Q$  on the left side, and $2_Q$ on the right, as depicted in figure \ref{fig2}. By symmetry, we have the relation
\begin{equation}
    w^\pm_{1_Q} = w^\mp_{2_Q}.
\end{equation}
The exact expression for the generalised entropy as a function of the QES locations can be written using the exact formulae for CFT entanglement entropy for free fermions \cite{Casini:2009sr, Hollowood:2021wkw} by labelling the six points  $\{2_L, 1_L,1_Q, 2_Q, 1_R, 2_R\}$ along the Cauchy slice as $\mu=1,2,\ldots 6$ from left to right,
\bea
S_I(A_L\cup A_R)&&=2S_0+ \frac{\pi c}{6\beta k}\sum_{\mu=3,4}\frac{1-w_\mu^+w_\mu^-}{1+w_\mu^+ w_\mu^-}-\frac{c}{12}\sum_{\mu\neq 3,4}\log w_\mu^+ w_\mu^- \label{Sisland}\\\nonumber
&&-\frac{c}{6}\sum_{\mu<\nu}(-1)^{\mu-\nu}\log\left[-(w_\mu^+-w_\nu^+)(w_\mu^--w_\nu^-)\right]-\frac{c}{6}\sum_{\mu=3,4}\log(1+w_\mu^+ w_\mu^-)\,.
\eea
The first two terms on the right-hand side arise from the dilaton evaluated at the QESs, the third and final terms originate from conformal factor contributions at endpoints in the bath and the QESs in the gravitating region. In what follows, we will work in the high temperature limit so that the extremal entropy $S_0$ is negligible in comparison to the thermal entropy of the black hole.
In general this is quite a complicated expression to extremise with respect to the QES locations $w_{1_Q}^\pm$ or $w_{2_Q}^\pm$. However, in the high temperature limit \eqref{highT}, and with $a< t<b$, the coordinates satisfy,
\be \label{near_horizon}
|w_{2_R}^\pm|, |w_{1_R}^+|\gg1 \,.
\ee
In addition, the QES is always expected to be close to the horizon in the semiclassical limit, and so we may reliably assume, say, for the right QES location,  
\be
|w^-_{2_Q}| \ll |w^+_{2_Q}|\,,\qquad  |w^+_{2_Q}w^-_{2_Q}| \ll 1\,,
\ee
which we can verify {\it a posteriori}. Finally, the ordering of points along the Cauchy slice in the high temperature limit implies, 
\be
w^+_{2_R} \gg w^+_{1_R}\gg w_{2Q}^+ \,.
\ee
With these assumptions (consistent with the semiclassical limit) in place,  extremising the generalised entropy with respect to the QES positions leads to the following equations, %assuming that $|w^-_{2_Q}| \ll |w^+_{2_Q}| \ll w^+_{1_R}, w^+_{2_R}$, and %$|w^-_{2Q}| \ll |w^-_{2R}|$
\begin{equation}\label{QESsolutionJT}
    \begin{split}
        &\frac{s+1}{s}\,w^+_{2_Q} = \frac{1}{w^-_{2_Q} - w^-_{1_R}}, \\
        &\frac{s+1}{s}\,w^-_{2_Q} = \frac{1}{w^+_{2_Q}} - \frac{1}{w^+_{2_Q} - w^+_{1_L}} + \frac{1}{w^+_{2_Q} - w^+_{2_L}} - \frac{1}{w^+_{1_R}}.
    \end{split}
\end{equation}
For the moment we retain $s$ in the numerator of the left-hand side, but eventually we want to consider $s\ll 1$. The equations for the QES  \eqref{QESsolutionJT} can be solved readily, and yield two sets of roots, one of which is physically relevant\footnote{The other root is negative and must be discarded.} and takes on a compact form (after using the condition $e^{4\pi a/\beta}\gg s/(s+1)$),
\bea
   w^+_{2_Q} = \sqrt{e^{2\pi(b+a-2t)/\beta}\,\sinh^2\tfrac\pi\beta(b-a)+\frac{2s}{s+1}e^{\pi(b-a)/\beta}\sinh\tfrac\pi\beta(b-a)}\hspace{0.6in}&&
   \nonumber\\\label{wQESplus}
   - e^{\pi(b+a-2t)/\beta}\cosh\tfrac\pi\beta(b-a)\,.&&
   %\\
%&&\Big[\sqrt{e^{-2t}(e^b - e^a)^2 + 4\frac{s}{s+1}\frac{e^a(e^b - e^a)}{e^{2a} - s/(s+1)}}- e^{-t}(e^b + e^a)\Big]/2.
\eea
This expression interpolates between early and late times within the time interval $a<t<b$ during which the entanglement dip is expected to occur,
\be
\label{qesregimes}
w_{2_Q}^+ \approx \begin{cases}
\frac{s}{s+1} \,e^{2\pi(t-a)/\beta}\qquad\qquad a< t \lesssim \frac{a+b}{2}+\frac12\,t_{\rm scr}\,,\\\\
\sqrt{\frac{s}{s+1}} \,e^{\pi(b-a)/\beta}\qquad\qquad \frac{a+b}{2} +\frac12 \,t_{\rm scr}\lesssim t< b\,.
\end{cases}
\ee
At this point, it is useful to make contact with the nomenclature introduced in \cite{Hollowood:2021wkw} where, ignoring $t_{\rm scr}$ for the moment, the following relevant temporal regimes were identified for the separate cases $b> 3a$:
\bea
&& ({\rm II}) =\{a< t< \tfrac12(b-a)\}\qquad ({\rm III}) =\{\tfrac12(b-a)< t< \tfrac12(a+b)\}\\\nonumber\\\nonumber
&&({\rm IV}) = \{\tfrac12(b+a)< t< b\}\,,
\eea
and $b<3a$:
\bea
&& ({\rm III}) =\{a< t< \tfrac12(a+b)\}\qquad({\rm IV}) = \{\tfrac12(b+a)< t< b\}\,.
\eea
For small $s$, ignoring $s$ in the denominator, the first line of \eqref{qesregimes} precisely matches the QES location in \cite{Hollowood:2021wkw} that covers the regimes (II) and (III) (across both cases above), whilst the second line matches the solution in regime (IV).\footnote{Furthermore, in the IR limit, the eternal AdS$_2$ black hole coupled to Minkowski baths can be argued to be  dual to two copies of (free) CFT with boundary (BCFT) prepared in the thermofield double state. The entanglement entropy of two intervals in the BCFT system  follows from a four-point correlator of twist fields. The different temporal regimes identified are controlled by different channels dominating the 4-point function. In the language of \cite{Hollowood:2021wkw}  the QES solutions \label{regimes} capture the entanglement evolution governed by the BCFT channels $(3)$ and $(4)$ which yield lightcone singularities in free CFTs.}  It is easy to check that the condition $|w_{2_Q}^-|\ll |w_{2_Q}^+|$ is automatically satisfied in these regimes, while the requirement that $|w_{2_Q}^+w_{2_Q}^-|\ll 1$ needs $s\ll1 $.
%For simplicity, we absorb $\beta/2\pi$ into the definition of $t$, $a$, and $b$. %As we are only interested in studying the dip region, we restrict ourselves to $a < t < b$.

Since the late time entropy in the no-island saddle is constant and equal to the thermal entropy, it is natural to consider the difference, 
\be
\Delta S \equiv S_I(A_L \cup A_R) - S_\varnothing (A_L \cup A_R)\,.
\ee
For late times this is essentially minus the mutual information. 
The exact expression for the no-island value of the entropy is 
\bea
S_\varnothing (A_L \cup A_R)&&=\label{Snoisland} \\\nonumber
-\frac{c}{12}&&\sum_{\mu}\log w_\mu^+ w_\mu^-
-\frac{c}{6}\sum_{\mu<\nu}(-1)^{\mu-\nu}\log\left[-(w_\mu^+-w_\nu^+)(w_\mu^--w_\nu^-)\right]\,,
\eea
where the sums run only over the 4 endpoints in the bath $\{2_L, 1_L, 1_R, 2_R\}$ labelled by $\mu, \nu=1,2,5,6$.
Then, 
%comparing the expressions for the island $S_I(A_L \cup A_R)$ and no-island $S_\varnothing(A_L \cup A_R)$ saddles, we define $\Delta S \equiv S_I(A_L \cup A_R) - S_\varnothing (A_L \cup A_R)$, 
the non-negligible contributions to $\Delta S$ can be compactly identified as,\footnote{As indicated previously, for simplicity we ignore the extremal entropy $S_0$ in the high temperature limit.}
%\footnote{Refer to Eqs. (4.8) and (4.12) of \cite{Hollowood:2021wkw} for the exact expressions of $S_\varnothing(A_L \cup A_R)$ and $S_I(A_L \cup A_R)$, respectively.}
%\begin{equation}
%\begin{split}
%    \Delta S &= \frac{1}{s}\frac{1 - w^+_{2Q}w^-_{2Q}}{1 + w^+_{2Q}w^-_{2Q}} - 2 \log(1 + w^+_{2Q}w^-_{2Q}) + 2\log(w^+_{2Q} - w^-_{2Q}) \\
%    &- 2\log(w^+_{2R} - w^+_{2Q})(w^-_{2Q} - w^-_{2R}) + 2\log(w^+_{2R} - w^-_{2Q})(w^+_{2Q} - w^-_{2R})\\
%    &- 2\log(w^+_{1R} - w^-_{2Q})(w^+_{2Q} - w^-_{1R}) + 2\log(w^+_{1R} - w^+_{2Q})(w^-_{2Q} - w^-_{1R}).
%\end{split}
%\end{equation}
%\begin{equation}
%    \Delta S \approx \frac{1}{s} + 2\log w^+_{2Q} + 2\log\frac{w^+_{2Q} - w^-_{2R}}{w^+_{2Q} - w^-_{1R}} + 2\log\frac{w^-_{2Q} - w^-_{1R}}{w^-_{2Q} - w^-_{2R}}.
%\end{equation}
\bea\label{DeltaS1stVersion}
    &&\Delta S = \\\nonumber
    && \frac{c}{6}\left[ \frac{1}{s}(1 - 2w^+_{2_Q}w^-_{2_Q}) + 2\log\frac{w^+_{2_Q} - w^-_{2_R}}{w^+_{2_Q} - w^-_{1_R}} + 2\log\frac{w^+_{2_Q}}{1 + w^+_{2_Q}(1 + 1/s)(w^-_{1_R} - w^-_{2_R})}\right].
\eea
In the high temperature limit and deep within the interval $a<t< b$, we can use $|w_{2_Q}^+|\gg |w_{1_R}^-|$  and $|w_{2_R}^-|\gg 1$ to write,
\bea\label{LocalMinimum}
   && \Delta S \approx\\\nonumber  
   &&\frac{c}{6}\left[\frac{1}{s}\left(\frac{1-s}{1+s} +2w^+_{2_Q}e^{{2\pi}(a - t)/\beta}\right) -2\log(1 + 1/s) + 2\log\left(e^{{2\pi}(t - b)/\beta} + \frac{1}{w^+_{2_Q}}\right)\right]\,.
\eea
To determine the critical size of the interval below which the entanglement dip disappears, we must find the minimum value of $\Delta S$ as a function of time and require that $\Delta S_{\rm min}\geq 0$.
Expecting a minimum near the midpoint of the interval $t=(a+b)/2$ we find that using {\em either} the early {\em or } the late time approximations for $w_{2_Q}^+$ in \eqref{regimes}, in the small $s$ limit, yields the same location for the minimum of eq.\eqref{LocalMinimum},
%In the regime of interest, we have $e^{2a} \gg s/(s+1)$, then $w^+_{2_Q}$ can be cast as
%\begin{equation}\label{w+, JT}
%    w^+_{2_Q} \approx \frac{s}{s+1}e^{t - a}.
%\end{equation}
% - \frac{s}{s+1}e^{t-b} - e^{-t+a}
\begin{equation}\label{SpikeShift}
    %e^{2\pi (t - b)/\beta} - (1 + 1/s)e^{-2\pi(t - a)/\beta} = 0 %\implies 
    { t_{\rm min} \approx \frac{a+b}{2} + \frac{\beta}{4\pi}\log\frac{1}{s}\,.}
\end{equation}
Although this result is at the edge of the validity of either the early or late time approximation for $w_{2Q}^+$, we find that it closely matches the results from the complete expressions \eqref{Sisland}, \eqref{wQESplus} and \eqref{Snoisland}.
We see above that the dip time is shifted from the midpoint of the interval by an amount fixed by the scrambling time of the black hole. Substituting $t_{\rm}$ in Eq. (\ref{LocalMinimum}), we obtain the value of $\Delta S_{\rm min}$,
\be
    \Delta S_{\text{min}} = \frac{c}{6}\left(- \frac{2\pi}{\beta }L+ \frac{1}{s} - \log\frac{1}{s} + {\cal O}(s^0)\right)\,,
    %\\
   % &{\color{blue}\Delta S_{\text{min}} = \frac{c}{6}\left( \frac{1}{s} - \log\frac{1}{s} - \frac{2\pi}{\beta }L +\mathcal{O}(1)\right)}\,
\ee
where $L = b - a$. Therefore the entanglement dip will vanish when $\Delta S_{\rm min} \geq0$ which in turn implies 
\be\label{CriticalCondition}
  L\leq L_{\rm crit}= \frac{\beta}{2\pi}\left( \frac{1}{s} -\log \frac{1}{s} + {\cal O}(s^0)\right)\,.
\ee
The critical length can also be re-expressed in terms of the scrambling time,
\be\label{lcritjt}
L_{\rm crit} \approx \frac{\beta}{2\pi}\,  \exp\left(\tfrac{2\pi}{\beta} t_{\rm scr}\right)- t_{\rm scr}\,.
\ee
This illustrates that in order to see long-distance correlations in the Hawking radiation, we must consider length/time scales that are at least exponential in scrambling time.
\subsection{Mutual information}
\begin{figure}[ht]
            \begin{center}
    \includegraphics[width=0.5\linewidth]{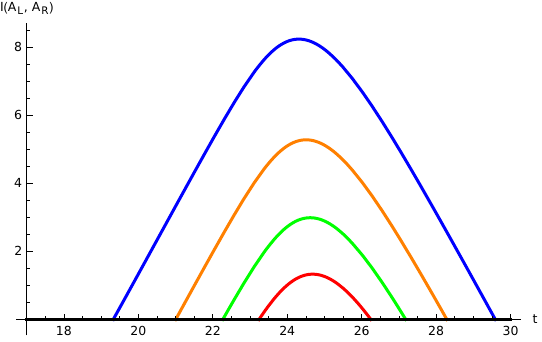}
            \caption{\small Mutual information $I(A_L,A_R)$ as a function of time $t$ for $a = 16$, $b= 30$, and $\beta = 2\pi$. The late-time spike is displayed for $s = 0.15$ (blue), $s = 0.1$ (orange), $s = 0.08$ (green), and $s = 0.07$ (red). For $s = 0.06$, scrambling effects completely dominate the correlation spreading process, and there is no spike. The black line corresponds to vanishing mutual information for values of $s\leq s_{\text{crit}}\approx 0.06$, where the Hawking or no-island saddle dominates at all times.}
            \label{fig3}
            \end{center}
\end{figure}
A dip in entanglement entropy corresponds to a spike in the mutual information $I(A_L,A_R)$, indicating that the correlations spread across spacetime as if they are being carried away by free quasiparticles \cite{Asplund:2015eha}. We now evaluate $I(A_L,A_R)$ using the exact expressions \eqref{EntropySingleInterval} together with \eqref{Snoisland} and \eqref{Sisland} for the no-island and island entropies, i.e.  $S_\varnothing(A_L \cup A_R)$ and  $S_I(A_L \cup A_R)$  respectively. 

In the following, we will first focus on the ``small interval" scenario in which the interval length is smaller than the distance  of the mid-point of the interval from the boundary:
\be
R_c\equiv \frac{b+a}{2}\,,\qquad L < R_c\implies b< 3a\,.
\ee
This regime, as noted in \cite{Hollowood:2021wkw}, is when the dip appears in isolation, well after the entropy of entanglement has plateaued. It is the only effect of the island saddle which results in a  spike or blip in the mutual information. We will treat separately the case $L > R_c$ or  $b>3a$ in the next subsection, since it reveals a second critical  regime for scrambling.

Our results show that entanglement memory is progressively lost as we decrease the parameter $s$, i.e., with increasing scrambling time. Figure \ref{fig3} shows that the spike in the mutual information, originating from the island contribution, is lower and narrower as $s$ is decreased for a fixed value of the interval length $L=14$ and $\beta=2\pi$.\footnote{Numerically small values of $\beta$ are problematic as they lead to very large exponents in our expressions, so we do not go deep into the high temperature limit.}
Interestingly, if we solve eq. \eqref{CriticalCondition} for $s$, with $L=14$ and $\beta = 2\pi$, we find  $s =0.059$ for the critical value of $s$ at which correlations should vanish, agreeing with the behaviour of the mutual information observed in figure \ref{fig3}. For any $s$ smaller $s\lesssim 0.06$, the correlations are completely scrambled. 

\begin{figure}[h]
            \centering
            \includegraphics[width=0.5\linewidth]{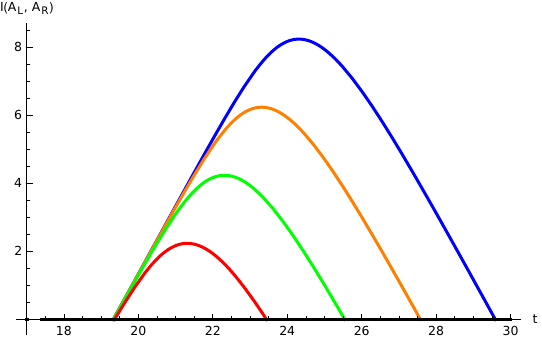}
            \caption{Mutual information $I(A_L, A_R)$ as a function of time $t$ for $s = 0.15$, $a = 16$, and $\beta = 2\pi$. We show the late-time spike for $b = 30$ (blue), $b = 28$ (orange), $b = 26$ (green), and $b = 24$ (red). The spike disappears when we set $b = 22$, and we note that $L_{\rm crit} \approx 6$ as dictated by eq.\eqref{CriticalCondition}. The black line corresponds to the mutual information for values of $L\leq L_{\text{crit}}$, where the island saddle is subdominant at all times.}
            \label{fig4}
\end{figure}

Additionally, %as predicted by eq.\eqref{SpikeShift}, the peak of the spike is slightly shifted to the right of the interval midpoint with decreasing $s$. 
we can find when the entanglement revivals start and end, considering corrections due to scrambling. Explicitly, the endpoints of the spike turn out to be
\begin{equation}\label{JTendpoints}
    t_{\text{in}}= a+\frac{1}{2k}+\mathcal{O}(\beta), \qquad\qquad t_{\text{fin}}= b-\frac{1}{2k}+\frac{\beta}{2\pi}\left(\log{\frac{1}{s}}+\mathcal{O}(s^0)\right).
\end{equation}
Therefore, although the peak time deviates from $t = (a+b)/2$ as we vary $s$, eq. \eqref{SpikeShift} shows that the spike remains symmetric up to high-temperature corrections about its peak at $t_{\rm min}= (t_{\rm in} + t_{\rm fin})/2$. Note that, in this regime, $t_{\text{in}}$ only depends on $a$ while $t_{\text{fin}}$ only depends on $b$. Thus, for instance, varying $b$ has no effect on $t_{\text{in}}$, as evident from the curves in figure \ref{fig4}. 

\begin{figure}[ht]
            \centering
            \includegraphics[width=0.5\linewidth]{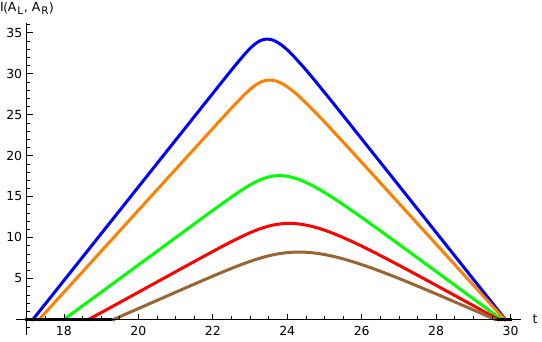}
            \caption{Mutual information $I(A_L, A_R)$ as a function of time for $s = 0.15$, $a = 16$, and $b = 30$ for different temperatures, $\beta/2\pi = 0.35$ (blue), $\beta/2\pi = 0.4$ (orange), $\beta/2\pi = 0.6$ (green), $\beta/2\pi = 0.8$ (red), and $\beta/2\pi = 1$ (brown). We only display the region where entanglement revivals occur. The black line represents the portion of time where the mutual information  is dominated by the Hawking saddle.}
            \label{fig5}
\end{figure}

Next, fixing $s$ and $\beta$, we investigate how mutual information responds to variations of $L$. We do this by fixing the endpoint $a$ and decreasing $b$.  In figure \ref{fig4}, we plot $I(A_L,A_R)$ after the initial correlations between the intervals are lost, focusing only on the entanglement revivals. For $L$ greater than the critical value $L_{\rm crit}$ determined by eq. \eqref{CriticalCondition}, the spikes start at the same time determined by the location of endpoint $a$, although their maximum point is shifted to the left as $b$ decreases. The spikes preserve their shape when $s$ and $L$ are varied, but this is not the case when we vary $\beta$ for fixed $s$ and $L$ as shown in figure \ref{fig5}. Decreasing $\beta$, whilst keeping other parameters fixed, we recover the sharp symmetric profile observed in \cite{Hollowood:2021wkw} in the high temperature, large interval limit. 

Note that the parameter $s$ depends on both $\beta$ and $k$, so we can keep it constant by scaling $\beta$ and the JT parameter $k$ whilst keeping their product fixed, and ensuring that $k \ll 1$ (recalling that we have set the AdS radius to unity) \cite{Almheiri:2019qdq}. For fixed $L$ and $s$, eq. \eqref{CriticalCondition} shows that the late-time spikes can be flattened by decreasing the temperature. This can be understood as progressively reducing the  correlations  shared by the intervals in the TFD state, making it easier for the black hole to scramble them.
\subsubsection{Additional critical scrambling for $b>3a$ case}\label{b>3aJT}
\begin{figure}[ht]
            \begin{center}
    \includegraphics[width=0.6\linewidth]{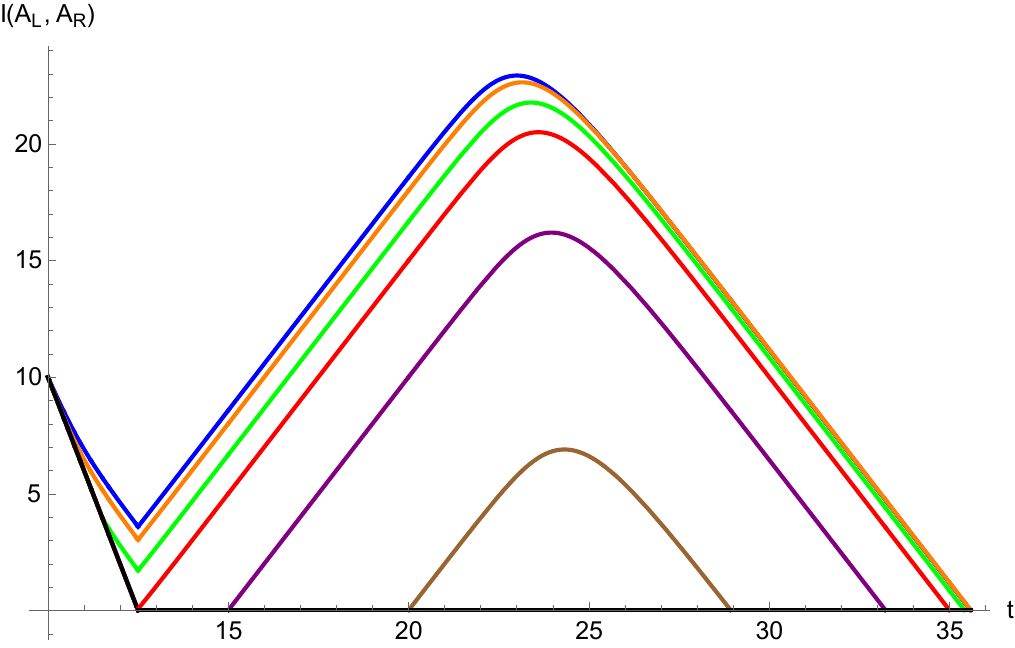}
            \caption{\small Mutual information $I(A_L,A_R)$ as a function of time $t$ for $a = 10$, $b= 35$, and $\beta = 2\pi$. The evaluation of $I(A_L,A_R)$ is displayed for $s = 0.7$ (blue), $s = 0.5$ (orange), $s = 0.3$ (green), $s = 0.2$ (red), $s = 0.1$ (purple), $s = 0.05$ (brown) and $s = 0.035$ (black). The red curve corresponds to $s_{\text{crit}}^{(\text{II})} = 0.2$ and the black line corresponds to the mutual information entirely determined by the Hawking/no-island saddle.}
            \label{fig6new}
            \end{center}
\end{figure}
In \cite{Hollowood:2021wkw} it was shown that for symmetric intervals satisfying $b>3a$, or $L> R_c$, there is an additional regime (II),  in which the island saddle starts dominating over the Hawking saddle {\em before} the entanglement plateau is reached, and whose effect is to  slow down the early-time entanglement entropy growth. This regime was considered in \cite{Hollowood:2021wkw} in the limit $L\gg t_{\text{scr}}$, and extends for $a<t<\frac{b-a}{2}$, following which, for  $\frac{b-a}{2}<t<b$ (regime (III))the island saddle produces an entanglement dip exactly as in the case $b<3a$ analysed previously, with minimum at $t=\frac{a+b}{2}$.\footnote{For context, the beginning of the entanglement dip falls in regime (III) and its subsequent rise after the minimum at $t_{\rm min}=\frac{a+b}{2}$ in regime (IV) \cite{Hollowood:2021wkw}.}

The analytic results for the QES in section \ref{critical_length_analysis} are valid for $a<t<b$ and for both  the cases $b > 3a$ and $b<3a$.  In the large interval situation  with $b>3a$, the time  $t_{\rm in}\simeq a+\frac{1}{2k}$ becomes the  time at which the island saddle starts dominating over the Hawking one in the regime (II) so the mutual information starts falling a bit slower than in the no-island saddle. A striking result is that, taking into account scrambling effects, the regime (II) starts later with respect to the BCFT prediction, and this regime itself  shrinks as scrambling time increases. In particular, there is a critical parameter $s_{\text{crit}}^{(\text{II})}$ below which the regime (II) disappears entirely in favour of the early-time entropy growth regime (I), and this critical point  can be found by equating $t_{\rm in}$ with the starting point of regime (III), i.e.
\begin{equation}
    a+\frac{1}{2k}=\frac{b-a}{2}\implies s_{\text{crit}}^{(\text{II})}= \frac{\beta}{2\pi}\,\frac{1}{b-3a}\,.
\end{equation}
In terms of the interval length $L$ and the distance of the interval midpoint from the boundary $R_c$, this condition can be re-expressed as 
\be
s_{\rm crit}^{(\rm II)}=\frac{\beta}{4\pi} \frac{1}{L-R_c}\,,
\ee
where $L$ is necessarily larger than $R_c$.
In figure \ref{fig6new}, we plot the mutual information $I(A_L,A_R)$ for a fixed choice of interval satisfying $b>3a$ and for several values of the scrambling parameter $s$: as expected, for $s<s_{\text{crit}}^{(\text{II})}$, the early linear decrease of mutual information is entirely due to the Hawking saddle.

\section{Entanglement revivals in the RST model}
\label{sec3}
We now turn to a discussion of finite interval effects in the RST model. The RST model is a soluble 2d dilaton gravity model coupled to a large-$N$ number of free fields and is a variant of the CGHS model \cite{Callan:1992rs}. It is an interesting toy model for studying Hawking evaporation in asymptotically flat spacetime. The generalised entropy formalism and the island paradigm, accompanied by a Page curve can be shown to arise from replica wormhole computations of the fine-grained entropy of entanglement of Hawking modes \cite{Hartman:2020swn} in this background. One of the features of this model is that the geometry has a boundary, much like the origin of space in radial coordinates, where one imposes reflecting boundary conditions for matter fields propagating on the background \cite{Fiola:1994ir}. The calculation of the von Neumann entropy of massless fields on such a background proceeds similarly to BCFT in two dimensions. Below we review some basic features of the RST model, before moving to an analysis of the finite interval problem for the single-sided evaporating black hole and the two-sided eternal black hole, both of which will exhibit entanglement dips for distinct reasons.

\subsection{Brief review of RST model}
The classical action of the RST model \cite{Russo:1992ax, Fiola:1994ir} is given by
\begin{equation}\label{classical_RST}
    S_{\text{cl}} = \frac{1}{2\pi}\int d^2 w \sqrt{-g} \left[e^{-2\phi}(R+4(\nabla \phi)^2+4\lambda^{2}) 
-\frac{N}{24}  \phi R-\sum_{k=1}^{N}i\bar\psi_k\slashed{\nabla} 
\psi_{k}\right],
\end{equation}
where $\phi$ is the dilaton field and $\{\psi_{k}\}$ are $N$ matter fields which we take to be massless Dirac fermions. The matter fields, in principle,  can be replaced by any CFT with central charge $N$. The action \eqref{classical_RST} without the linear $\phi R$ term yields the CGHS model \cite{Callan:1992rs} --  two-dimensional dilaton gravity theory describing the low-energy physics governing the radial modes of a near extremal four-dimensional magnetically charged black hole. 
The length scale $\lambda^{-1}$ accounts for the magnetic charge of the four-dimensional black hole, and from now on we will set $\lambda=1$. A consistent semiclassical approximation can be implemented by taking the large-$N$ limit, specifically $N\to +\infty$ with $Ne^{2\phi}$ fixed. At leading order in $1/N$, it turns out that the quantum fluctuations of the dilaton and the metric are negligible, and we need only to include the effect of the conformal anomaly due to matter fields  i.e.
\begin{equation} 
    S_{\rm anom} = -\frac{N}{96\pi} \int d^2 w \sqrt{-g}R\,\Box^{-1}R\,.
\end{equation}
The linear $\phi R$ counterterm, distinguishing the RST from the CGHS action, has been added to make the model analytically tractable in the large-$N$ limit, and it has been seen numerically \cite{Lowe:1992ed, Piran:1993tq} that this fine-tuned counterterm does not change the physics of the model qualitatively.

Going to the conformal gauge $ds^2 = -e^{2\rho}dw^+ dw^-$, with $w^{\pm} = w^0\pm w^1$ and $R_{+-}=-2\partial_+\partial_-\rho$, the large-$N$ effective action becomes
%\begin{align}\label{rsteff}
%S_{\rm eff}=\frac{1}{2\pi}\int d^2 x\, \left[e^{-2\phi}\left(4\partial_{+}\rho\partial_{-}\rho-8 \partial_{+}\phi\partial_{-}\phi+2e^{2\rho}\right) +\frac{N}{6}(\rho-\phi) \partial_+\partial_-\rho +\sum_{k=1}^N\partial_+ f_k\partial_- f_k\right]\,.
%\end{align}
\bea
\label{rsteff}
S_{\rm eff}
&&
= \frac{1}{2\pi} \int d^2 w \,\left[
e^{-2\phi}\left(4\,\partial_{+}\rho\,\partial_{-}\rho
- 8\,\partial_{+}\phi\,\partial_{-}\phi
+ 2 e^{2\rho}\right) +\right.
\nonumber \\&&
\left.\hspace{1.1in}+ \,\frac{N}{6}(\rho - \phi)\,\partial_{+}\partial_{-}\rho + \sum_{k=1}^{N}i\left( \psi_{k+}\partial_{-} \psi_{k+}+ \psi_{k-}\partial_{+} \psi_{k-}\right)
\right] \,,
\eea
where $\psi_{k\pm}$ are left- and right-moving fermions. 
At this point, it is convenient to perform the following field redefinitions, 
\begin{equation}\label{omegachi}
\Omega = {\frac{12}{N}}e^{-2\phi}+\frac{\phi}{2}-\frac{1}{4} \log {\frac{48}{N}}\,,\qquad \chi = {\frac{12}{N}}e^{-2\phi}+\rho-\frac{\phi}{2}+\frac{1}{4} \log {\frac{3}{N}}\,,
\end{equation}
where the field $\Omega$ plays the role of the area of the two-sphere of the four-dimensional black hole, therefore it is the RST gravity counterpart of the JT gravity dilaton.

In terms of these fields the action takes the form, 
\bea\label{bilca}
&&S_{\rm eff}=\frac{N}{12\pi}\int d^2 w \,\left(\partial_+ \Omega\partial_- \Omega\, -\,\partial_+\chi \partial_-\chi \,+\, e^{2\chi-2\Omega}\,+\right.\\\nonumber
&&\left.\hspace{2.6in}+\frac{6}{N}\sum_{k=1}^Ni\left( \psi_{k+}\partial_{-} \psi_{k+}+ \psi_{k-}\partial_{+} \psi_{k-}\right)\right)\,,
\eea
which scales with an overall factor of $N$ in the large-$N$ limit, with $\Omega$ and $\chi$ held fixed. 

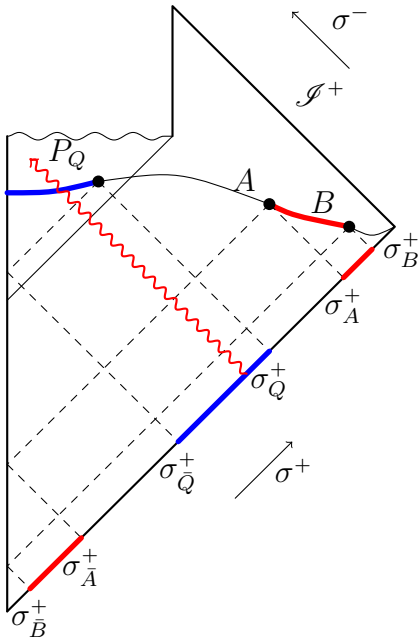
\begin{figure}[h]
\centering
\begin{comment}
\begin{minipage}{0.49\textwidth}
\centering
    \begin{tikzpicture}[scale=1.5, line cap=round, line join=round]
  %\draw[thick]
    %(0,0) -- (0,4.2) -- (1.45,4.2) -- (1.45,5.35) -- (3.4,3.4) -- cycle;
    \draw[thick] (0,0) -- (0,4.2);
    \draw[thick] (1.45,4.2) -- (1.45,5.35) -- (3.4,3.4) -- (0,0);
  \draw[decorate, decoration={snake, amplitude=0.5mm, segment length=5mm}]
    (0,4.2) -- (1.45,4.2);

  \fill (0.8,3.8) circle (1.5pt) node[above left=0.1pt] {$P_Q$};
  \fill (2.3,3.6) circle (1.5pt) node[above left=1pt] {$A$};
  \fill (3.0,3.4) circle (1.5pt) node[above left=1pt] {$B$};

  \draw[black] (1.45,4.2) -- (0,2.75);
  \draw[blue, thick] (0,3.7) -- (0.8,3.8);
  \draw[red, thick] (2.3,3.6) -- (3.0,3.4);
  \draw[black, thick] (0.8,3.8)--(2.3,3.6);
  \draw[black, thick] (3.0,3.4)--(3.4,3.4);
  \node[right] at (3.4,3.4) {$i^{0}$};
  \node[above right] at (2.425,4.375) {${\mathscr I}^+$};

  \draw[->] (2,1.5) -- (2.5,2) node[midway, right] {$\sigma^{+}$};
  \draw[->] (3,4.8) -- (2.5,5.3) node[midway, above right] {$\sigma^{-}$};
  
\end{tikzpicture}
\end{minipage}
\end{comment}
\centering
\begin{tikzpicture}[scale=1.5, line cap=round, line join=round]
  %\draw[thick]
    %(0,0) -- (0,4.2) -- (1.45,4.2) -- (1.45,5.35) -- (3.4,3.4) -- cycle;
    \draw[thick] (0,0) -- (0,4.2);
    \draw[thick] (1.45,4.2) -- (1.45,5.35) -- (3.4,3.4) -- (0,0);
  \draw[decorate, decoration={snake, amplitude=0.5mm, segment length=5mm}]
    (0,4.2) -- (1.45,4.2);

    \draw[red, line width = 2pt] (2.3,3.6) ..controls (2.5,3.5) .. (3.0,3.4);
  \draw[blue, line width = 2pt] (0,3.7) .. controls (0.4,3.7) ..  (0.8,3.8);
  \draw[black] (0.8,3.8) .. controls (1.5,3.9) ..  (2.3,3.6);
  \draw[black] (3.0,3.4) .. controls (3.2,3.3) ..  (3.4,3.4);

  \fill (0.8,3.8) circle (1.5pt) node[above left=0.1pt] {$P_Q$};
  \fill (2.3,3.6) circle (1.5pt) node[above left=1pt] {$A$};
  \fill (3.0,3.4) circle (1.5pt) node[above left=1pt] {$B$};

  \draw[dashed] (2.3,3.6) -- (0,1.3);
  \draw[dashed] (3.0,3.4) -- (0,0.4);
  \draw[dashed] (0.8,3.8) -- (0,3);

  \draw[dashed] (2.3,3.6) -- (2.95,2.95);
  \draw[dashed] (3.0,3.4) -- (3.2,3.2);
  \draw[dashed] (0.8,3.8) -- (2.3,2.3);

  \draw[dashed] (0,1.3) -- (0.65,0.65);
  \draw[dashed] (0,0.4) -- (0.2,0.2);
  \draw[dashed] (0,3) -- (1.5,1.5);

  \draw[red, line width = 2pt] (0.2,0.2) -- (0.65,0.65);
  \draw[red, line width = 2pt] (2.95,2.95) -- (3.2,3.2);
  \draw[blue, line width = 2pt] (1.5,1.5) -- (2.3,2.3);

  \node[below] at (0.2,0.2) {$\sigma_{\bar{B}}^{+}$};
  \node[below] at (0.65,0.65) {$\sigma_{\bar{A}}^{+}$};
  \node[below] at (1.5,1.5) {$\sigma_{\bar{Q}}^{+}$};
  \node[below] at (2.3,2.3) {$\sigma_{Q}^{+}$};
  \node[below] at (2.95,2.95) {$\sigma_{A}^{+}$};
  \node[right] at (3.2,3.2) {$\sigma_{B}^{+}$};
  \node[above right] at (2.425,4.375) {${\mathscr I}^+$};

  \draw[->] (2,1) -- (2.5,1.5) node[midway, right] {$\sigma^{+}$};
  \draw[->] (3,4.8) -- (2.5,5.3) node[midway, above right] {$\sigma^{-}$};

  \draw[black] (1.45,4.2) -- (0,2.75);
  \draw[red, decorate,thick,decoration={snake,segment length=2mm,amplitude=0.5mm},->] (2.1,2.1) -- (0.2,4.0); 
  \end{tikzpicture}
\caption{The Penrose diagram depicting an evaporating black hole, produced by a shockwave (red wavy line) insertion at $\sigma^+=0$, as described by the solution (\ref{evaporatingBHsol}) of the RST model. The region $R$ and its projections onto ${\mathscr I}^-$ are shown in red, whereas the island and its corresponding null projection are shown in blue.}
\label{fig6}
\end{figure}

The field $\Omega$ has a minimum as a function of $\phi$, where $\Omega_{\rm min}=1/4$. A smaller value of $\Omega$ would require a complex dilaton, which would be problematic,\footnote{A complex dilaton would give a negative coefficient of the Einstein-Hilbert term, thus a violation of unitarity in the semiclassical description.} implying the well-known condition
\begin{equation}
    \Omega\geq \frac{1}{4}\,.
\end{equation}
This condition yields a boundary “singularity" in the geometry, specified by the curve $\Omega(w^{+},w^{-})=\frac{1}{4}$. For the single-sided evaporating black hole, this singularity consists of a timelike and a spacelike portion (see figure \ref{fig6}). The former can naturally be viewed as the origin of radial coordinates in a higher-dimensional setting and where one simply imposes reflecting boundary conditions on the $N$ matter fields $\{\psi_{k\pm}\}$ upon dimensional reduction. Therefore, we are dealing with a boundary conformal field theory (BCFT) for the matter fields. 
The spacelike portion of the curve $\Omega(w^{+},w^{-})=\frac{1}{4}$ is naturally interpreted as the black hole singularity behind the horizon. Consistent with this picture, for the eternal black hole the extended spacetime has no timelike boundary, and $\Omega(w^{+},w^{-})=\frac{1}{4}$ yields the spacelike singularities cloaked behind the past and future horizons as depicted in figure \ref{fig7}. 

The action \eqref{bilca} has a residual gauge symmetry, which can be fixed by making the following choice 
\begin{equation}
    \chi=\Omega\,,
\end{equation}
which turns out to describe the geometry in Kruskal coordinates. In this gauge, the dilaton and the generalised area can be expressed in terms of the conformal factor as, 
\begin{align}
    &\phi=\rho_{K}-\frac{1}{2}\log\frac{N}{12}\,,\\
    &\Omega= e^{-2\rho_{K}}+\frac{1}{2}(\rho_{K}-\log2)\,.\label{OmegaRST}
\end{align}
The generalised area is determined by the equation of motion
\begin{equation}\label{eom}
\partial_+\partial_- \Omega= -1\,,
\end{equation}
together with a constraint following from the semiclassical Einstein equations,
\begin{equation}
\partial_\pm^2 \Omega = \,-\langle T_{\pm\pm} \rangle\,.\label{constraint}
\end{equation}
where on the RHS we have the expectation value of the rescaled energy-momentum tensor 
\begin{equation}
    T_{\pm\pm}=\frac{6}{N}\sum_{k=1}^{N}i\,\psi_{k\pm}\partial_{\pm}\psi_{k\pm}\,,
\end{equation}
normal ordered with respect to the Kruskal vacuum state, containing no positive-frequency modes with respect to the Kruskal time.
\subsection{Evaporating black hole}
Before analysing the entanglement entropy of a finite-size radiation subregion $R$ at ${\mathscr I}^+$ in the case of an evaporating black hole, let us review the main features of the solution itself.  

An evaporating black hole can be created by sending in a null pulse of energy $M$ at $w^{+}=1$, and the corresponding geometry, following from eqs. \eqref{eom} and \eqref{constraint}, is described by,\footnote{We follow the conventions adopted in literature on the RST model, e.g., \cite{Fiola:1994ir} corresponding to a late time inverse Hawking temperature of $\beta =2\pi$. In these units, the mass $M$ of the black hole is dimensionless.}
\begin{equation}\label{evaporatingBHsol}
\Omega(w^+, w^-) = - w^+ w^- -\frac14\log (-4w^+ w^-)- M(w^+-1)\Theta(w^+-1)\,.
\end{equation}
The edge of spacetime, given by the curve $\Omega=\frac{1}{4}$ as explained above, is timelike behind the shockwave. On this portion we impose reflecting boundary conditions on the matter fields. It becomes spacelike after impact with the shockwave -- this portion is interpreted as the black hole singularity, after which the boundary becomes timelike again at the endpoint when the black hole has evaporated completely.

The evaporating black hole solution \eqref{evaporatingBHsol} has an apparent horizon, defined as the boundary of the trapped points, explicitly
\begin{equation}
    \partial_{+}\Omega=0 \implies w^{+}(w^{-}+M)=-\frac{1}{4}\,, \qquad w^{+}>1\,,
\end{equation}
and the end of evaporation is given by the intersection of the apparent horizon with the black hole singularity,
\begin{equation}
    w_{\rm EP}^{-}=\frac{M}{e^{-4M}-1}\,, \qquad w_{\rm EP}^{+}=\frac{1}{4M}(e^{4M}-1)\,.
\end{equation}
The event horizon will be defined by $w^{-}=w^{-}_{\rm EP}$.

The ``in" vacuum for the fields is determined by the background at ${\mathscr I}^-$,
\begin{equation}
e^{2\rho_K}\big|_{w^- \to -\infty} = -\frac{1}{w^+w^-} +\frac{M}{(w^-)^2}\frac{(w^+ -1)}{(w^{+})^{2}} + O\left(\frac{1}{(w^{-})^{3}}\right)\,,
\end{equation}
which corresponds to the linear dilaton ``Minkowski" vacuum \cite{Russo:1992ax, Fiola:1994ir}. The ``out" vacuum, on the other hand, is fixed by asymptotics at ${\mathscr I}^+$,
\begin{equation}
e^{2\rho_K}\big|_{w^+ \to \infty} = -\frac{1}{w^+(w^-+M)} +O\left(\frac{1}{(w^{+})^{2}}\right)\,.
\end{equation}
The natural outgoing Minkowski coordinates, using the same formalism as in \cite{Hartman:2020swn}, are 
\begin{equation}
e^{\tilde{\sigma}^+}= w^+\,,\qquad -e^{-\tilde{\sigma}^-}= w^-+M\,,
\end{equation}
related to the incoming coordinates,
\begin{equation}w^\pm = \pm e^{\pm \sigma^\pm}\,.
\end{equation}
Equivalently,
\begin{equation}
\tilde{\sigma}^+=\sigma^+\,,\qquad \tilde{\sigma}^- = \sigma^--\log(1-Me^{\sigma^-})\,.
\end{equation}
The endpoint of evaporation in these two sets of coordinates is,
\bea
&&\sigma^+_{\rm EP}=\tilde{\sigma}^+_{\rm EP}= 4M -\log 4M + \log(1-e^{-4M})\,,\nonumber\\\label{endopoint2}\\\nonumber
&&\sigma^-_{\rm EP}= -\log M + \log(1-e^{-4M})\,,\qquad \tilde{\sigma}^-_{\rm EP}= 4M+\sigma^-_{\rm EP}\,.
\eea
\subsubsection{Finite interval radiation entropy}
The overall state is pure on any everywhere-spacelike or null slice stretching from spatial infinity to the timelike portion of the boundary singularity $\Omega=\frac{1}{4}$, at which we impose reflecting boundary conditions on the matter fields $\{\psi_{k}\}$.

The analogous problem, analysed above for the case of a black hole in JT gravity in thermal equilibrium with finite-size radiation intervals in the bath, can be rephrased in RST gravity's evaporating black hole scenario by considering a finite-size radiation interval at ${\mathscr I}^+$.

As a preliminary step, let us consider an interval $R\equiv\left[P_{A},P_{B}\right]$ on a slice (prior to the future light-cone of the evaporation endpoint) of the type described above. The quantum state on $R$ is a density matrix $\rho_{R}$ obtained by tracing over all the degrees of freedom of the slice outside $R$. With reflecting boundary conditions, the single-interval entanglement entropy in a BCFT can be viewed as the two-interval entanglement entropy of a chiral CFT on the plane. This was also extensively explained in \cite{Fiola:1994ir}. Thus the entanglement entropy $S_{\text{ent}}(R)=-\text{Tr}(\rho_{R}\log{\rho_{R}})$ is actually the entropy of {\em two disjoint intervals}, one corresponding to the direct null projection of $R$ on ${\mathscr I}^-$ and the other obtained by considering the reflection of  $R$ across the timelike boundary, projected on ${\mathscr I}^-$. 

On ${\mathscr I}^-$  the quantum fields are in their vacuum state, therefore for the entanglement entropy calculation, we can employ the free fermion results found in \cite{Casini:2009sr}.\footnote{In the presence of a shockwave, as in our case, it is sufficient to consider a coherent state of matter fields, and the same formulas apply.}
The quantum extremal prescription imposes that, in order to compute the generalised entropy of $R$, we need to take into account an island region as well, which we will denote as $I$, extending from the timelike boundary singularity to a point $P_{Q}$ on the same Cauchy slice as the finite radiation interval. Therefore, our central object will be 
\begin{equation}
    S(R)=\text{min}\,\text{ext}_{P_{Q}}\left[S_{\text{grav}}(P_{Q})+S_{\text{\rm CFT}}(I\cup R)\,\right]\,,
\end{equation}
where 
\begin{align}
    %&S_{\text{gen}}(I\cup R)= S_{\text{grav}}(P_{Q})+S_{\text{ent}}(I\cup R)\,,\\
    &S_{\text{grav}}(P_{Q})=\frac{N}{6}\left(e^{-2\rho_{K}(P_{Q})}-\frac{\rho_{K}(P_{Q})}{2}\right) +\frac{N}{24}(\log{4}-1)+\frac{N}{6}\log{\epsilon_{UV}}\,,\label{Bekenstein-HawkingRST}
\end{align}
such that $S_{\text{grav}}$ is the area term  in RST gravity \cite{Fiola:1994ir, Hartman:2020swn},
% as a consequence of the purity of the state on the whole Cauchy slice
and the CFT entanglement entropy of $I\cup R$ corresponds to the entanglement entropy of {\em three} disjoint intervals of finite size in a chiral (left-moving) CFT, namely
\begin{align}
    S_{\text{CFT}}(I\cup R)&=\frac{N}{6}\left(\rho_{\text{in}}(P_{Q})+\rho_{\text{in}}(P_{A})+\rho_{\text{in}}(P_{B}))\right) -\frac{N}{2}\log(\epsilon_{UV})\,\notag\\
    &+\frac{N}{6}\left[\sum_{i,j}\log{|u_{i}-v_{j}|}-\sum_{i<j}\log{|u_{i}-u_{j}|}-\sum_{i<j}\log{|v_{i}-v_{j}|}\right]
\end{align}
with
\bea
    [u_{1},v_{1}]=[\sigma^{+}_{\bar{B}},\sigma^{+}_{\bar{A}}]\,,
    \qquad [u_{2},v_{2}]=[\sigma^{+}_{\bar{Q}},\sigma^{+}_{Q}]\,,\qquad
    [u_{3},v_{3}]=[\sigma^{+}_{A},\sigma^{+}_{B}]\,.
\eea
Here we note that the incoming Minkowski coordinate of the null ray which reaches a point $P$, after reflecting off the boundary at $\Omega=\frac14$, is given by
\begin{equation}
    \sigma_{\bar{P}}^{+}=\sigma_{P}^{-}-\log4\,.
\end{equation}
Since we want the observer to collect the outgoing radiation at ${\mathscr I}^+$, we can move the Cauchy slice towards ${\mathscr I}^+$ by taking $\sigma^+_{A,B}\to +\infty$. The entanglement entropy of the outgoing radiation then only depends on the coordinates $\sigma^-_{A,B}$ , or equivalently the reflected coordinates $\sigma^+_{\bar A,\bar B}$ on ${\mathscr I}^-$,   of the endpoints $P_A$ and $P_B$. This yields the following expression for the CFT entanglement entropy of $I\cup R$,
\begin{equation}\label{EE-IR}
    S_{\text{\rm CFT}}(I\cup R)=\frac{N}{6}\log{\left[\frac{(\sigma_{Q}^{+}-\sigma_{\bar{B}}^{+})(\sigma_{\bar{A}}^{+}-\sigma_{\bar{B}}^{+})(\sigma_{\bar{Q}}^{+}-\sigma_{\bar{A}}^{+})(\sigma_{Q}^{+}-\sigma_{\bar{Q}}^{+})}{\epsilon_{\rm UV}^{2}e^{-\rho_{\text{in}}(P_{A})-\rho_{\text{in}}(P_{B})-\rho_{\text{in}}(P_{Q})}(\sigma_{\bar{Q}}^{+}-\sigma_{\bar{B}}^{+})(\sigma_{Q}^{+}-\sigma_{\bar{A}}^{+})}\right]}\,,
\end{equation}
where one factor of  $\epsilon_{UV}$ factor is accounted for by taking  $\sigma^+_{A}-\sigma^+_{B}\simeq \epsilon_{\rm UV}$.

The area term \eqref{Bekenstein-HawkingRST}, evaluated at the QES $P_{Q}$, can be equivalently expressed as  
\begin{equation*}
    S_{\text{grav}}(P_{Q})=\frac{N}{6}\left[\Omega(P_{Q})-\rho_{K}(P_{Q})+\log 2-\frac{1}{4}+\log\epsilon_{\rm UV}\right]\,,
\end{equation*}
by  using eq.\eqref{OmegaRST}.

Putting all together, the generalised entropy of $I\cup R$ takes the form
\begin{align}\label{generalizedentropy-RST}
    S_{\text{gen}}(I\cup R)&=S_{\text{grav}}(P_{Q})+S_{\text{CFT}}(I\cup R)\\
    &=\frac{N}{6}\left[\Omega(P_{Q})-\frac{1}{4}+\frac{1}{2}(\sigma_{Q}^{+}-\sigma_{\bar{Q}}^{+})+\log(\sigma_{Q}^{+}-\sigma_{\bar{Q}}^{+})+\log(\sigma_{Q}^{+}-\sigma_{\bar{B}}^{+})\;+\right.\notag\\
    &\left.\quad+\log(\sigma_{\bar{Q}}^{+}-\sigma_{\bar{A}}^{+})-\log(\sigma_{\bar{Q}}^{+}-\sigma_{\bar{B}}^{+})-\log(\sigma_{Q}^{+}-\sigma_{\bar{A}}^{+})\right]\;+\notag\\
    &\quad+\frac{N}{6}\log\left[\frac{\sigma_{\bar{A}}^{+}-\sigma_{\bar{B}}^{+}}{\epsilon_{UV}\sqrt{\left(1-4Me^{\sigma_{\bar{A}}^{+}}\right)\left(1-4Me^{\sigma_{\bar{B}}^{+}}\right)}}\right]\notag\,,
\end{align}
where we used 
\begin{align}
    &\rho_{\text{in}}=\rho_{K}+\frac{1}{2}(\sigma^{+}-\sigma^{-})\,,\\
    &\rho_{\text{in}}(P_{i})=\log\left(\frac{1}{\sqrt{1-4Me^{\sigma_{\bar{i}}^{+}}}}\right)\,,\qquad\qquad i=A,B\,.
\end{align}
The last line of eq. \eqref{generalizedentropy-RST} is the standard CFT result, the coarse-grained entanglement entropy, of the radiation interval $R$. The extremisation of the generalised entropy with respect to QES coordinates $\sigma_{\bar{Q}}^{+}$ and $\sigma_{Q}^{+}$ yields, respectively,
\begin{align}
    &e^{(\sigma_{Q}^{+}-\sigma_{\bar{Q}}^{+})}+1+\frac{4}{\sigma_{Q}^{+}-\sigma_{\bar{Q}}^{+}}-\frac{4}{\sigma_{\bar{Q}}^{+}-\sigma_{\bar{A}}^{+}}+\frac{4}{\sigma_{\bar{Q}}^{+}-\sigma_{\bar{B}}^{+}}=0\label{extremality1}\,,\\
    &e^{(\sigma_{Q}^{+}-\sigma_{\bar{Q}}^{+})}+1-4Me^{\sigma_{Q}^{+}}+\frac{4}{\sigma_{Q}^{+}-\sigma_{\bar{Q}}^{+}}-\frac{4}{\sigma_{Q}^{+}-\sigma_{\bar{A}}^{+}}+\frac{4}{\sigma_{Q}^{+}-\sigma_{\bar{B}}^{+}}=0\,.\label{extremality2}
\end{align}
Usually one is interested in the entropy of {\em all} radiation emitted by the black hole up to some fixed time in which case the radiation interval is effectively semi-infinite stretching from spacelike infinity to some point $P$ on $\mathcal{I}^{+}$, and the Page curve is obtained as a function of the outgoing Minkowski coordinate of the endpoint $\Tilde{\sigma}_{P}^{-}$ \cite{Hartman:2020swn}. In our case, it is natural to pick the outgoing coordinate of the furthest endpoint $\Tilde{\sigma}_{A}^{-}$ as the temporal dial, while the size of the interval is determined  by the difference 
\be
L_{AB}\equiv \Tilde{\sigma}_{A}^{-}-\Tilde{\sigma}_{B}^{-}\,.
\ee
We will keep $L_{AB}$ fixed whilst $\tilde\sigma_{A,B}^-$ both grow linearly with outgoing Minkowski time.

For large enough interval size $L_{AB}\gg1$, the CFT or coarse-grained  entropy for $R$ first grows linearly and then saturates at late times,  
\be\label{CFTentropy}
S_{\rm CFT}(R)\simeq\begin{cases}
 \frac{N}{12}\,\left(\tilde{\sigma}_{A}^{-}+\log{M}+2\log{L_{AB}}\right)\qquad &\tilde{\sigma}_{A}^{-}\ll L_{AB}-\log{M}\,,\\\\
 \frac{N}{12}\,L_{AB}\qquad &\tilde{\sigma}_{A}^{-}\gtrsim L_{AB}-\log{M}\,.
\end{cases}
\ee
This result from the no-island or Hawking saddle must be thought of as a balancing act between modes entering and modes leaving a finite sliding window: the endpoint $A$ continually captures fresh Hawking modes, while the endpoint $B$ lets older modes escape. At early times, 
as more modes enter the interval than escape, the Hawking entropy increases until the plateau time $\tilde{\sigma}_{A}^{-}\simeq L_{AB}-\log{M}$, at which point the entry and exit rates equalize, and the entropy of the interval saturates at the thermal value.\footnote{From the one-point function for the stress tensor, it can be readily deduced that the evaporating RST black hole has an effective time dependent temperature that quickly settles to a constant value $\beta^{-1}\approx 1/2\pi$. The thermal entropy of a large interval of size $L$ is then simply $\pi N L/6\beta$, matching \eqref{CFTentropy}.}

\begin{figure}[ht]
            \centering
            \includegraphics[width=0.495\linewidth]{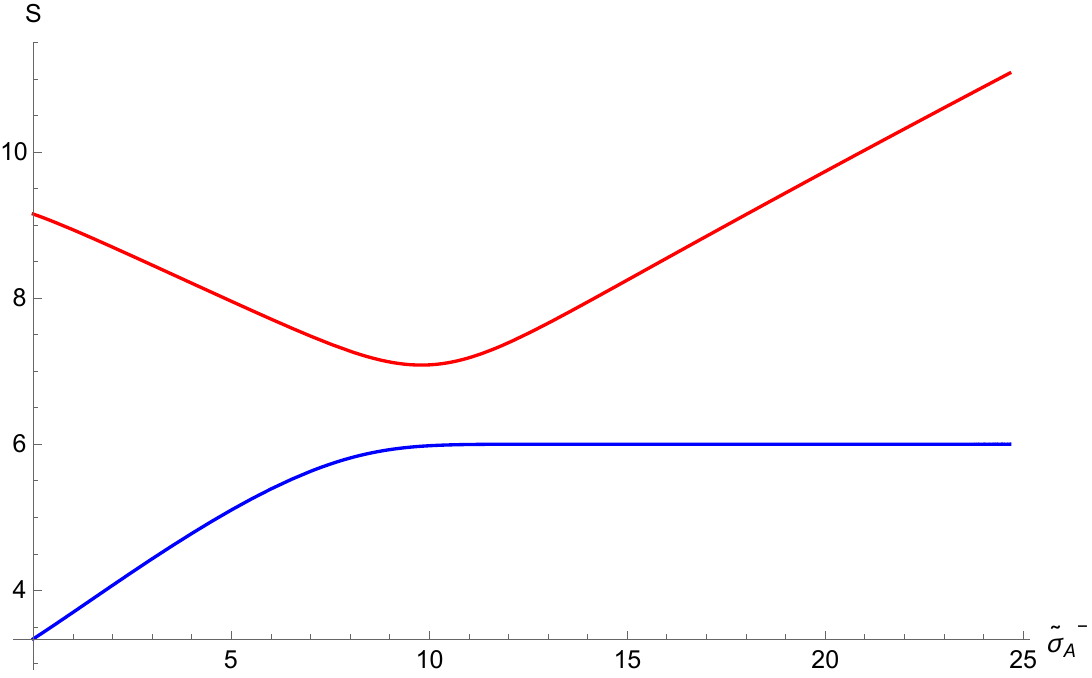}
            \includegraphics[width=0.495\linewidth]{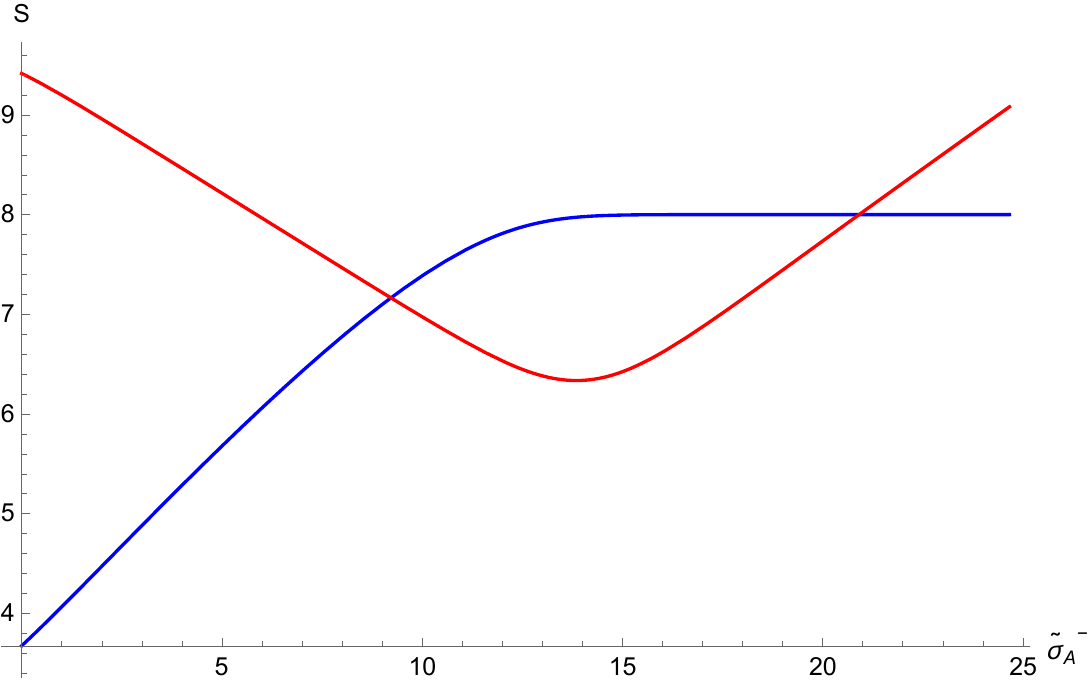}
            \caption{Time evolution of the generalised entropy due to the Hawking (blue) and island (red) saddles, with $M=7$ and $L_{AB}=12$ (left), $16$ (right). An entanglement dip appears for interval sizes above a critical value.}
            \label{fig7}
\end{figure}

Numerical solution of the QES equations \eqref{extremality1} and \eqref{extremality2} in the regime $M\gg1$ shows that the QES rapidly settles close to the event horizon, located at $\sigma_{\rm EP}^{-}\simeq -\log{M}$. Taking the difference of  eqs. \eqref{extremality1} and \eqref{extremality2}, we obtain,
\begin{equation}\label{sigmaQ+ev}
    e^{\sigma_{Q}^{+}}=\frac{1}{M}\left(\frac{1}{\sigma_{Q}^{-}-\Tilde{\sigma}_{A}^{-}+\log{(1+Me^{\Tilde{\sigma}_{A}^{-}})}}-\frac{1}{\sigma_{Q}^{-}-\Tilde{\sigma}_{A}^{-}+\log{(e^{L_{AB}}+Me^{\Tilde{\sigma}_{A}^{-}})}}\right)\,.
\end{equation} 
where we suppress the analogue $\sigma_{Q}^{+}$ terms on the RHS, since they are subleading {\em a posteriori}.\\
Eqs. \eqref{sigmaQ+ev} and  \eqref{extremality1}, at leading order then imply,
\begin{comment}
\begin{equation}
    4\left(\frac{e^{-\sigma_{Q}^{-}}}{M}-1\right)\left(\frac{1}{\sigma_{Q}^{-}-\Tilde{\sigma}_{A}^{-}+\log{(1+Me^{\Tilde{\sigma}_{A}^{-}})}}-\frac{1}{\sigma_{Q}^{-}-\Tilde{\sigma}_{A}^{-}+\log{(e^{L_{AB}}+Me^{\Tilde{\sigma}_{A}^{-}})}}\right)=0\,,
\end{equation}
\end{comment}
\begin{equation}
    4e^{\sigma_{Q}^{+}}\left(e^{-\sigma_{Q}^{-}}-M\right)=0\,,
\end{equation}
which yields $\sigma_{Q}^{-}=-\log{M}$, agreeing with  the numerical solutions. This solution for $\sigma_Q^-$ along with  \eqref{sigmaQ+ev} determines $\sigma_{Q}^+$
\begin{equation}\label{sigmaqplus}
    \sigma_{Q}^{+}=\Tilde{\sigma}_{A}^{-}-L_{AB}+\log{(e^{L_{AB}}-1)}+\mathcal{O}\left(\frac{e^{-\tilde{\sigma}_A^-}}{M}\right)\,.
    %\left(\frac{e^{-\Tilde{\sigma}_{A}^{-}+L_{AB}}}{(1-e^{L_{AB}})M}\right)\,.
\end{equation}
In figure \ref{fig7} we display the time evolution of the generalised entropy evaluated on the no-island (blue) and island (red)  saddles, for two different values of $L_{AB}$, up to the endpoint of evaporation given by  $\Tilde{\sigma}_{A,\text{fin}}^{-}=4M-\log{M}+L_{AB}-\log{(e^{L_{AB}}-1)}$, at which the QES hits the endpoint of evaporation \eqref{endopoint2}. As we keep increasing the interval size, it turns out that the island saddle becomes more relevant, and for a larger time range. In the strict $L_{AB}\to +\infty$ limit, we recover the Page curve of a semi-infinite interval with $S_{\text{gen}}(\Tilde{\sigma}_{A,\text{fin}}^{-})=S_{\text{gen}}(\Tilde{\sigma}_{A,\text{in}}^{-})$ obtained in \cite{Hartman:2020swn, Gautason:2020tmk}.

To find the critical interval length below which the island saddle is always subdominant and no entanglement dip appears, we first note  that the difference between the generalised entropy evaluated on the island and the Hawking saddles takes the following form, 
\bea
    &&\frac{6\Delta S}{N}\simeq \\\nonumber
    &&M-\frac{3}{4}(\log{M}+\Tilde{\sigma}_{A}^{-})+\log{\left[\frac{L_{AB}+\log{4}+\log{\left(1+Me^{-L_{AB}+\Tilde{\sigma}_{A}^{-}}\right)}}{L_{AB}-\Tilde{\sigma}_{A}^{-}-\log{M}+\log{\left(1+Me^{-L_{AB}+\Tilde{\sigma}_{A}^{-}}\right)}}\right]}\,,
\eea
where we have used $L_{AB}\gg1$. %since we are interested in the dip regime and the critical transition at which the island saddles becomes always subdominant. 
We will see that this is justified {\em a posteriori} as the critical value for $L_{AB}$ is $\mathcal{O}(M)$. The minimum of $\Delta S$ is  located at $\Tilde{\sigma}_{A}^{-}\simeq L_{AB}-\log{M}$, with value,
\begin{equation}\label{DeltaSmin_RST_ev}
    %\frac{6\Delta S_{\text{min}}}{N}\simeq M-\frac{3}{4}\left(L_{AB}+\log\frac{3}{2}\right)-\log{\left(\log{\frac{5}{3}}\right)}+\log{L_{AB}}\,.
    \frac{6\Delta S_{\text{min}}}{N}= M-\frac{3}{4}L_{AB}+\log{L_{AB}}+\mathcal{O}(M^0)\,.
\end{equation}
The island and Hawking saddles give comparable entropy contributions for $L_{AB}\gtrsim M$ as expected.
%justifying the leading contributions in \eqref{DeltaSmin_RST_ev}.\\
In particular, the critical length above which a dip appears, satisfying $\Delta S_{\rm min}=0$, is 
\begin{equation}\label{criticalenRST}
    %L_{AB}^{\text{crit.}}\simeq -\frac{4}{3}W_{-1}\left(-\frac{3}{4}e^{-M}\right)\simeq \frac{4}{3}\left(M+\log{M}-\log\frac{4}{3}\right)+\mathcal{O}\left(\frac{\log{M}}{M}\right)\,.
    L_{AB}^{\text{crit}}\simeq -\frac{4}{3}W_{-1}\left(-\frac{3}{4}e^{-M}\right)= \frac{4}{3}\left(M+\log{M}\right)+\mathcal{O}\left(M^0\right),
\end{equation}
here $W_k(x)$ is the product logarithm function.\footnote{ It is implicitly defined through $W_k(x)e^{W_k(x)} = x$ and $k$ indicates choice of branch of the function.}

However, note that, in the RST model, the scrambling time and mass of the black hole are related via \cite{Fiola:1994ir}
\begin{equation}\label{scrambling, RST}
    t_{\text{scr}}=\frac{\beta}{2\pi}\log\frac{S_{\rm BH}}{N}=\log\frac{M}{6}\,,
\end{equation}
recalling that $\beta=2\pi$ and $S_{\rm BH}=\frac{NM}{6}$. Therefore in terms of scrambling time we can express the critical length as follows
\begin{equation}
    L_{AB}^{\text{crit}}= 8 e^{t_{\text{scr}}}+\frac{4}{3}t_{\text{scr}}+\mathcal{O}\left(1\right).
\end{equation}

In addition, we can also compute the Page time when the entanglement dip begins ($\Delta S =0$) for any fixed $L_{ AB}$, assuming $L_{AB}> L_{AB}^{\rm crit}$,
\begin{equation}
    \Tilde{\sigma}_{A,\text{Page}}^{-}\simeq L_{AB}-\log{M}+\frac{4}{3}W_{-1}\left(-\tfrac{3}{4} e^{M-\frac{3}{4}L_{AB}}L_{AB}\right)\,.
\end{equation}
The reality of the last term imposes the minimum or critical length requirement on $L_{AB}$. In the $L_{AB}\to +\infty$, we recover the Page time for the semi-infinite radiation interval \cite{Hartman:2020swn},
\begin{equation}
    \Tilde{\sigma}_{A,\text{Page}}^{-}(L_{AB}\to +\infty) \simeq \frac{4}{3}M-\log{M}\,.
\end{equation}

\subsubsection{Interpretation of dip and rise}
When the island saddle becomes dominant, the entanglement book-keeping  changes. The QES rapidly approaches the event horizon, and its outgoing coordinate  $\sigma_Q^+$, tracks the radiation window as per \eqref{sigmaqplus}. In this way, the island captures the purifying partners of a subset of the Hawking quanta contained in $R$, and the fine-grained entropy decreases.

However, for a finite interval, this purification is only temporary. As the coordinates $\tilde\sigma^-_{A,B}$ of the two endpoints describe a sliding interval of fied size, the radiation window keeps moving. Some  Hawking modes that were purified by the island eventually leave the interval through the endpoint $B$. Once they are no longer in $R$, the corresponding purification is lost from the entropy of $I\cup R$, and the island branch can rise again. This explains the second rise: it is a finite-window effect caused by already-purified Hawking modes escaping the radiation interval. 

Following the dip and the rise, generically the single interval entanglement entropy will saturate at its thermal value.
For $L_{AB}$ large enough, this loss through $B$ is postponed beyond the evaporation regime, and the second rise disappears, getting a Page-curve behaviour, analogous to the one of a semi-infinite radiation region.

On the other hand, if we consider increasingly small radiation intervals, there is a critical size dictated by eq.\eqref{criticalenRST} below which the interval never contains enough Hawking quanta for the partial purification to compensate the gravitational cost of including the island, i.e. the island saddle is never dominant and the radiation is maximally scrambled.

\begin{comment}
One can check numerically that in this scenario there is no dip for any value of $L_{AB}\equiv \sigma_{A}^{-}-\sigma_{B}^{-}$. \\
The von Neumann entropy initially increases due to the dominance of the semiclassical Hawking saddle. At the Page time, the island saddle takes over, causing the entropy curve to bend. Near the end of evaporation $\tilde{\sigma}^{-}\lesssim \tilde{\sigma}^{-}_{\text{EP}}\simeq4M-\log M$, the von Neumann entropy reaches a finite and non-zero value, due to the fact we are not tracing out the whole radiation system, up to spacelike infinity $i^{0}$.
\end{comment}

\subsection{Eternal RST black hole and entanglement revival}

\begin{figure}
    \centering
    \begin{tikzpicture}
                [scale=0.7]
% Top triangle (no top horizontal line)
\fill[gray!20] (0,6) -- (-3,3) -- (3,3) -- cycle;
\draw (0,6) -- (-3,3) -- (0,6);
\draw (0,6) -- (3,3);

% Bottom triangle (no bottom horizontal line)
\fill[gray!20] (0,-6) -- (-3,-3) -- (3,-3) -- cycle;
\draw (0,-6) -- (-3,-3) -- (0,-6);
\draw (0,-6) -- (3,-3);
%\draw[pink,fill=Plum!10!pink] (-3,-3) -- (3,-3) -- (3,3) -- (-3,3) -- cycle;
%
\draw[-] (-3,3) -- (-6,0) -- (-3,-3);
\draw[-]  (3,-3) -- (6,0) -- (3,3);
\draw[decorate, decoration={snake, amplitude=0.5mm, segment length=5mm}]
    (-3,-3) -- (3,-3);
\draw[decorate, decoration={snake, amplitude=0.5mm, segment length=5mm}]
    (-3,3) -- (3,3);
\draw[thick] (-3,-3) -- (3,3);
\draw[thick] (-3,3) -- (3,-3);
\filldraw[black] (3.25,1.5) circle (2pt);
\filldraw[black] (-3.25,1.5) circle (2pt);
\filldraw[black] (4.7,0.5) circle (2pt);
\filldraw[black] (-4.7,0.5) circle (2pt);
\filldraw[black] (-1.4,1.6) circle (2pt);
\filldraw[black] (1.4,1.6) circle (2pt);
%
%
%\node at (-4.7,0.2) {\footnotesize $2_L$};
%\node at (-3.25,1.2) {\footnotesize $1_L$};
%\node at (3.25,1.2) {\footnotesize $1_R$};
%\node at (4.7,0.2) {\footnotesize $2_R$};
\node at (0, 2.2) {\footnotesize $I$};
%
%\draw[dashed] (-6,0) to[out=-170,in=18] (-4.7,0.5);
\draw[thick, red] (-3.25,1.5) to[out=-170,in=18] (-4.7,0.5);
%\draw[dashed] (-3.25,1.5) to[out=0,in=190] (-1.6,1.6);
\draw[thick, blue] (-1.4, 1.6) to[out=10,in=162] (1.4, 1.6);
%\draw[dashed] (1.6, 1.6) to[out=0,in=190] (3.25,1.5);
\draw[thick, red] (3.25,1.5) to[out=-10,in=162] (4.7,0.5);
%\draw[dashed] (4.7,0.5) to[out=-10,in=162] (6,0);
%
\node at (3.7,0.6) {\footnotesize $A_R$};
\node at (-3.7,0.6) {\footnotesize $A_L$};
\node at (0,3.4) {\footnotesize $\Omega = 1/4$};
\node at (0,-3.6) {\footnotesize $\Omega = 1/4$};
\draw[->] (0,0) -- (0.6,0.6) node[right] {$w^{+}$};
\draw[->] (0,0) -- (-0.6,0.6) node[left] {$w^{-}$};
%\node at (3.3,3.3) {\footnotesize $w^+$};
%\node at (-3.3,3.3) {\footnotesize $w^-$};
%
%\draw[<->] (2.7,0.3) -- (3,0) -- (3.3,0.3);
%\node at (2.4,0.6) {\footnotesize $x^-$};
%\node at (3.6,0.6) {\footnotesize $x^+$};
%
%\draw[<->] (-0.6,0.6) -- (0,0) -- (0.6,0.6);
%\node at (-0.5,0.9) {\footnotesize $w^-$};
%\node at (0.6,0.9) {\footnotesize $w^+$};
%
%\begin{scope}[yshift=0cm]
%\draw[<->] (-2.7,-0.3) -- (-3,0) -- (-3.3,-0.3);
%\node at (-2.4,-0.6) {\footnotesize $x^-$};
%\node at (-3.6,-0.6) {\footnotesize $x^+$};
%\end{scope}
%
%\draw[<->] (0.3,1.2) -- (0,1.5) -- (0.3,1.8);
%\node at (0.6,0.9) {\footnotesize $x^-$};
%\node at (0.6,2.1) {\footnotesize $x^+$};
%
            \end{tikzpicture}
    \caption{Eternal black hole described by the solution \eqref{EternalBHRST} of the RST model. Now, the $\Omega =1/4$ curve is spacelike everywhere and represents the singularity behind the horizon, depicted by wavy lines.}
    \label{fig8}
\end{figure}
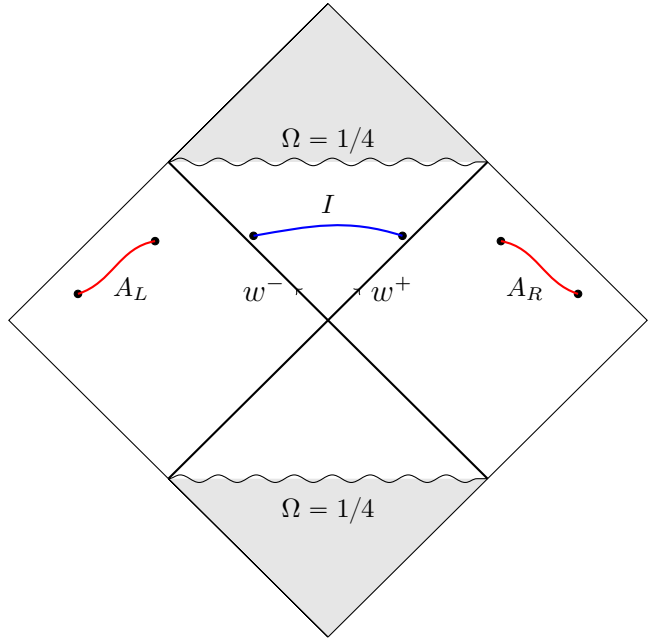

The solution for the eternal black hole in the RST model is given by
\begin{equation}\label{EternalBHRST}
    \Omega=-w^{+}w^{-}+M\,,
\end{equation}
satisfying Eqs. \eqref{eom} and \eqref{constraint} with $\langle T_{\pm\pm} \rangle=0$. The Kruskal coordinates are defined as in the AdS$_2$ case \eqref{Kruskal}, with the only difference being that the inverse temperature is fixed to $\beta=2\pi$.

Similarly to the JT gravity-plus-bath system, we now consider two symmetric intervals on ${\mathscr I}^+$, with $a,b\gg1$, as forming the radiation system. 
The calculation of the generalised entanglement entropy, including the island contribution, follows the exact same steps as in the JT case. The main differences arise in the area (Bekenstein-Hawking) contribution to the generalised entropy and the conformal factors at the endpoints of the intervals, respectively given as,
\begin{align}
    &S_{\text{grav}}=\frac{N}{6}\left(e^{-2\rho_{K}}-\frac{\rho_{K}}{2}\right)+\frac{N}{24}(\log4-1)+\frac{N}{6}\log{\epsilon_{\rm UV}}\,,\\
    &S_{\text{conf}}=\frac{N}{6}\rho_{K}-\frac{N}{6}\log{\epsilon_{\rm UV}}\,.
\end{align}
Given two symmetrically located QESs, the expression for the generalised entropy in the presence of an island is then,
\bea
S_{\text{gen}}(I\cup R)&&= \label{Sgen-RST}\\\nonumber
 && 
 \frac{N}{3}
 \left[\Omega(w_{2_Q}^\pm)-\frac{1}{4}-\log\frac{\epsilon_{\rm UV}^2}{2}+\rho_{K}(w_{1_R}^\pm)+\rho_{K}(w_{2_R}^\pm)\right]+S_{\rm CFT}^{\rm flat}(I\cup R)\,,
\eea
where $S_{\rm CFT}^{\rm flat}(I\cup R)$ is the flat space, vacuum entanglement entropy of $I\cup R$ \cite{Casini:2009sr}, identical to corresponding terms in the JT-gravity case \eqref{Sisland}.

%In the high-temperature/large-interval limit, $\Omega(w_{1_R}^\pm), \Omega(w_{2_R}^\pm) \gg1$ and 

%The generalised entropy \eqref{Sgen-RST} can be simplified as follows
%\begin{equation}
%    S_{\text{gen}}(I\cup R)=\frac{N}{3}\left[\Omega(w_{2_Q})+\log2-\frac{1}{4}-2\log\epsilon_{\rm UV}-\frac{1}{2}\log\Omega(w_{1_R})\Omega(w_{2_R})+(\dots)\right]\,,
%\end{equation}
%using
%\begin{equation}
%    \rho_{K}=2\Omega+\log2+\frac{1}{2}W_{-1}(-e^{-4\Omega})\simeq -\frac{1}{2}\log\Omega+\mathcal{O}\left(\frac{\log4\Omega}{\Omega}\right)\,,
%\end{equation}
%where the first equality is due to Eq. \eqref{OmegaRST}, while the approximation is justified by $\Omega \gg1$ at the extrema of radiation intervals.\\

The extremisation conditions with respect to the QES coordinates $w_{2_Q}^\pm$, in the high temperature/large interval limit, are the same as eq. \eqref{QESsolutionJT}, with the replacement $(s+1)/s\rightarrow1$. The coordinate $w_{2_Q}^+$ of the right  QES is,  
\be\label{regimesRST}
w_{2_Q}^+ \approx \begin{cases}
e^{t-a}\;\quad\qquad\qquad a< t \lesssim \frac{a+b}{2}\,,\\\\
e^{(b-a)/2} \qquad\qquad \frac{a+b}{2}\lesssim t< b\,.
\end{cases}
\ee
Restoring the temperature dependence, this is essentially the same as \eqref{regimes}. The difference between the island and no-island generalised entropies is,
\begin{align}
\frac{6\Delta S}{N} &\approx 2M -2w_{2_Q}^{+}w_{2_Q}^{-}+2\log2-\frac{1}{2} + 2\log\left(e^{t - b} + \frac{1}{w^+_{2_Q}}\right)\,.
%&\simeq 2M -2\left(1-e^{\frac{a+b}{2}-t}\right)+2\log2-\frac{1}{2}+2\log\left(e^{t-b}+e^{\frac{a-b}{2}}\right)\,,
\end{align}
%where in the last step we used Eq. \eqref{regimesRST}.\\
The location of the dip is  found by minimising  $\Delta S$ with respect to time, 
\begin{equation}\label{RSTendpoints}
    t_{\text{\rm min}}=\frac{a+b}{2} +\mathcal{O}(M^0)\,,
\end{equation}
and the effect disappears for interval length below a critical value,
\begin{equation}\label{critical-length-eternalRST}
    L_{\rm crit}=2M+\mathcal{O}(M^0)= 12 e^{t_{\text{scr}}}+\mathcal{O}(1)\,,
\end{equation}
where in the second equality we used the relation between mass and scrambling time in the RST model stated in eq. \eqref{scrambling, RST}. Interestingly, in contrast to \eqref{SpikeShift} and \eqref{lcritjt} for the AdS$_2$ case in JT-gravity, the location of the dip and critical length both do not display ${\cal O}(t_{\rm scr})$ corrections.\par
Below, we plot the results for the mutual information $I(A_L, A_R)$ of the left and right radiation intervals by keeping $L_{AB}$ fixed and varying the black hole mass $M$, and vice versa. As expected from our analytical result \eqref{critical-length-eternalRST}, entanglement memory progressively disappears upon either increasing the scrambling time or shrinking the interval length. The absence of $\mathcal{O}(t_{\text{scr}})$ corrections to the $t_{\rm min}$ makes the mutual information peak appear symmetric about the mid-point unlike the situation in JT gravity. Moreover, the endpoints are found to be
\begin{equation}\label{tintfinRST}
    t_{\text{in}}= a+M+\mathcal{O}(1), \qquad\qquad t_{\text{fin}}= b-M+\mathcal{O}(1)\,,
\end{equation}
again without $\mathcal{O}(t_{\text{scr}})$ corrections.

We focussed above on the case $b<3a$, while for the case $b>3a$, analogous results to the ones found in \ref{b>3aJT} follow: in particular the critical mass for the disappearance of regime (II) is,
\begin{equation}
    M_{\text{crit}}^{\text{(II)}}=\frac{b-3a}{2}\,.
\end{equation}
\begin{figure}[H]
            \centering
            \includegraphics[width=0.49\linewidth]{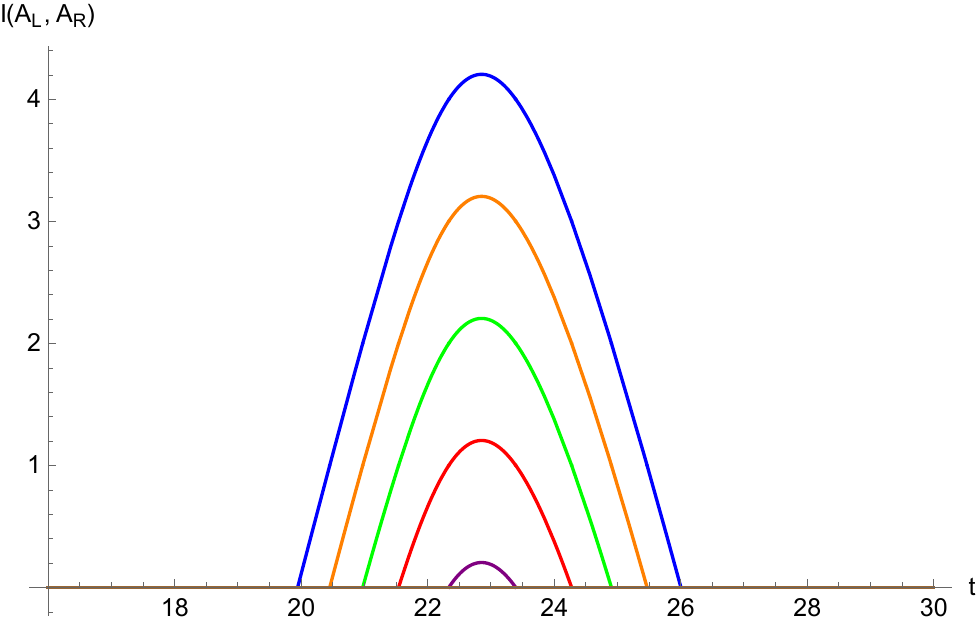}
            \includegraphics[width=0.49\linewidth]{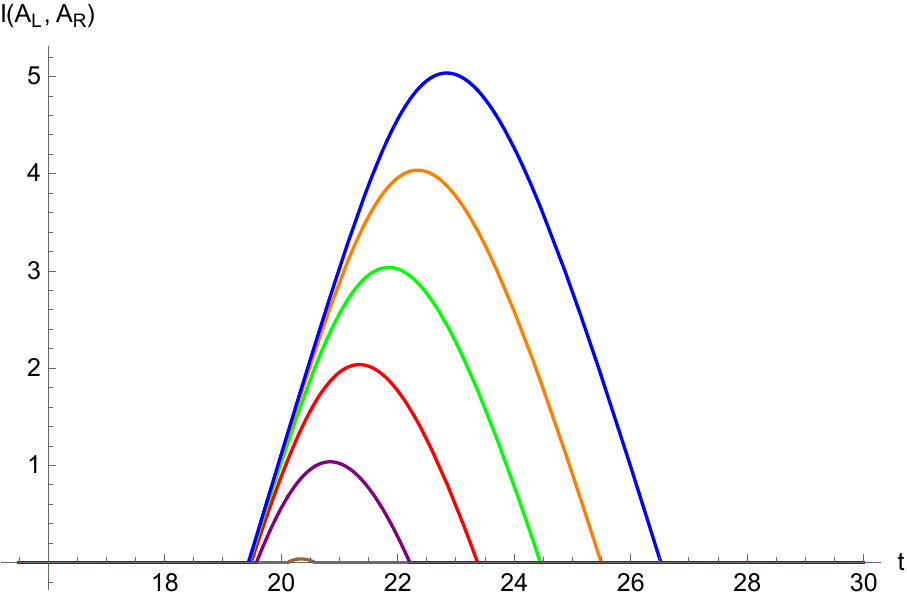}
            \caption{Depiction of the mutual information $I(A_L, A_R)$ after the entropic early-time growth regime. {\it Left}: the size of the intervals is fixed, with $a=16$ and $b=30$, and the black hole mass increases from the top curve in blue ($M=3.5$) to the bottom curve $(M=6)$ in steps of $0.5$. {\it Right}: the black hole mass and one endpoint of each of the two symmetric intervals is kept fixed, $M=3$ and $a=16$, while the size of the intervals is decreased, starting with $b=29$ up to $b=24$ in steps of $1$.}
            \label{fig9}
\end{figure}

\section{Conclusions and open questions}
\label{sec4}
We studied the entanglement evolution of finite, equal-size intervals in $2$d CFT prepared in the thermofield double state, viewing CFT degrees of freedom as radiation modes of evaporating black holes. The black holes under consideration are solutions of two well-known $2$d dilaton gravity theories: JT gravity and RST gravity. For the JT case, we considered an eternal black hole coupled to flat space baths  CFT$_L$ and CFT$_R$ in the high-temperature limit (in the TFD state). For the RST model, we analysed both an evaporating black hole with a finite radiation interval on $\mathscr{I}^+$ and the corresponding eternal black hole setup. In each case, we computed the fine-grained entropy and the mutual information, investigating how island-induced purification effects are modified by the interplay between the interval size and the scrambling time.

The general lesson from both these setups is that entanglement revival sets in for length scales larger than the exponential of the black hole scrambling time with  ${\cal O}(t_{\rm scr})$ corrections \eqref{crituniversal}. This is, of course, not unexpected as the Page time itself scales with the entropy of the black hole. The critical length scale, however, also provides a measure of the length scale above which long-range correlations in the radiation become detectable.

There are a few different questions of interest following on from our work. In the limit of zero scrambling time and high temperatures, the entanglement structure of two disjoint finite intervals $A_L$ and $A_R$ in JT gravity coupled to free Minkowski CFT baths, precisely matches the entanglement evolution of free BCFTs in the TFD state \cite{Hollowood:2021wkw}. In this situation the appearance of the dip can be explained through modes reflecting off the boundary in either copy, entering  $A_L$ and $A_R$ along with their respective purifiers in the TFD copy. It would be interesting to find the correct extension or modification of this geometric picture once the scrambling effects become relevant: in particular the arguments in \cite{Callebaut:2018nlq} suggest that the boundary behaves as if it has an effective thickness $\epsilon\sim \frac{1}{k}$. We would then expect rays meeting the boundary to undergo a delay and/or clipping effect, which could potentially explain our results for the endpoints of the dip (or mutual information spike) being shifted inwards by an amount $\frac{1}{2 k}$, which is actually exponential in scrambling time. This is the case in both JT and RST models seen in eqs. \eqref{JTendpoints} and \eqref{tintfinRST}.  Exploring this connection could also provide a route to incorporating the effects of finite scrambling time of the boundary theory on the modular Hamiltonian of the free BCFT \cite{Reyes:2021npy}. 

Our focus in this paper was on the free fermion CFT with scrambling induced purely by the black hole, or in the holographic dual description by a boundary SYK-like chaotic system (for a closely related setting see \cite{Fritzsch:2021bpm, Fritzsch:2023byj}).   For maximally scrambling holographic large-$c$ CFTs we know that entanglement revivals are absent with or without a black hole \cite{Asplund:2015eha, Asplund:2013zba, Balasubramanian:2021xcm}. In principle it would be more interesting to look at situations where the radiation system was neither free nor maximally scrambling, so the interplay between the scrambling times of the radiation system and black hole could be investigated in more detail \cite{Liu:2020gnp, Gu:2017njx, Chen:2017dbb}. This type of situation could potentially be explored by considering the SYK system coupled to interacting spin chains, for instance \cite{Chen:2020wiq}.

\acknowledgments 
The authors would like to thank Tim Hollowood for discussions. SPK would like to acknowledge support from STFC Consolidated Grant Award ST/X000648/1. NB is supported by the STFC DTP grant reference no. ST/Y509644/1. LBNB is supported by EPSRC Grant EP/W524694/1 and STFC Grant ST/Y509644/1.
\\
\\
{\small {\bf Open Access Statement} - For the purpose of open access, the authors have applied a Creative Commons Attribution (CC BY) licence to any Author Accepted Manuscript version arising. 

Data access statement: no new data were generated for this work.}


\newpage
\begin{thebibliography}{99}

%\cite{Hayden:2007cs}
\bibitem{Hayden:2007cs}
P.~Hayden and J.~Preskill,
``Black holes as mirrors: Quantum information in random subsystems,''
JHEP \textbf{09} (2007), 120
%doi:10.1088/1126-6708/2007/09/120
[\arXiv{0708.4025} [hep-th]].
%1266 citations counted in INSPIRE as of 26 Mar 2026

%\cite{Sekino:2008he}
\bibitem{Sekino:2008he}
Y.~Sekino and L.~Susskind,
``Fast Scramblers,''
JHEP \textbf{10} (2008), 065
%doi:10.1088/1126-6708/2008/10/065
[\arXiv{0808.2096} [hep-th]].
%1210 citations counted in INSPIRE as of 26 Mar 2026

%\cite{Shenker:2013pqa}
\bibitem{Shenker:2013pqa}
S.~H.~Shenker and D.~Stanford,
``Black holes and the butterfly effect,''
JHEP \textbf{03} (2014), 067
%doi:10.1007/JHEP03(2014)067
[\arXiv{1306.0622} [hep-th]].
%1526 citations counted in INSPIRE as of 26 Mar 2026

%\cite{Asplund:2015eha}
\bibitem{Asplund:2015eha}
C.~T.~Asplund, A.~Bernamonti, F.~Galli and T.~Hartman,
``Entanglement Scrambling in 2d Conformal Field Theory,''
JHEP \textbf{09} (2015), 110
%doi:10.1007/JHEP09(2015)110
\arXiv{1506.03772} [hep-th].
%159 citations counted in INSPIRE as of 31 Mar 2026


%\cite{Calabrese:2009qy}
\bibitem{Calabrese:2009qy}
P.~Calabrese and J.~Cardy,
``Entanglement entropy and conformal field theory,''
J. Phys. A \textbf{42} (2009), 504005
%doi:10.1088/1751-8113/42/50/504005
\arXiv{0905.4013} [cond-mat.stat-mech].
%1688 citations counted in INSPIRE as of 06 Apr 2026


%\cite{Asplund:2013zba}
\bibitem{Asplund:2013zba}
C.~T.~Asplund and A.~Bernamonti,
``Mutual information after a local quench in conformal field theory,''
Phys. Rev. D \textbf{89} (2014) no.6, 066015
%doi:10.1103/PhysRevD.89.066015
\arXiv{1311.4173} [hep-th].
%102 citations counted in INSPIRE as of 06 Apr 2026

%\cite{Leichenauer:2015xra}
\bibitem{Leichenauer:2015xra}
S.~Leichenauer and M.~Moosa,
``Entanglement Tsunami in (1+1)-Dimensions,''
Phys. Rev. D \textbf{92} (2015), 126004
%doi:10.1103/PhysRevD.92.126004
\arXiv{1505.04225} [hep-th].
%62 citations counted in INSPIRE as of 06 Apr 2026



%\cite{Jackiw:1984je}
\bibitem{Jackiw:1984je}
R.~Jackiw,
``Lower Dimensional Gravity,''
Nucl. Phys. B \textbf{252} (1985), 343-356
%doi:10.1016/0550-3213(85)90448-1
%1210 citations counted in INSPIRE as of 06 Apr 2026

%\cite{Teitelboim:1983ux}
\bibitem{Teitelboim:1983ux}
C.~Teitelboim,
``Gravitation and Hamiltonian Structure in Two Space-Time Dimensions,''
Phys. Lett. B \textbf{126} (1983), 41-45
%doi:10.1016/0370-2693(83)90012-6
%1166 citations counted in INSPIRE as of 31 Mar 2026

%\cite{Russo:1992ax}
\bibitem{Russo:1992ax}
J.~G.~Russo, L.~Susskind and L.~Thorlacius,
``The Endpoint of Hawking radiation,''
Phys. Rev. D \textbf{46} (1992), 3444-3449
%doi:10.1103/PhysRevD.46.3444
[\arXiv{hep-th/9206070} [hep-th]].
%347 citations counted in INSPIRE as of 06 Mar 2026

%\cite{Fiola:1994ir}
\bibitem{Fiola:1994ir}
T.~M.~Fiola, J.~Preskill, A.~Strominger and S.~P.~Trivedi,
``Black hole thermodynamics and information loss in two-dimensions,''
Phys. Rev. D \textbf{50}, 3987-4014 (1994)
%doi:10.1103/PhysRevD.50.3987
\arXiv{hep-th/9403137} [hep-th].
%226 citations counted in INSPIRE as of 09 Jan 2025

%\cite{Almheiri:2019yqk}
\bibitem{Almheiri:2019yqk}
A.~Almheiri, R.~Mahajan and J.~Maldacena,
``Islands outside the horizon,''
\arXiv{1910.11077} [hep-th].
%403 citations counted in INSPIRE as of 31 Mar 2026

%\cite{Almheiri:2019qdq}
\bibitem{Almheiri:2019qdq}
A.~Almheiri, T.~Hartman, J.~Maldacena, E.~Shaghoulian and A.~Tajdini,
``Replica Wormholes and the Entropy of Hawking Radiation,''
JHEP \textbf{05} (2020), 013
%doi:10.1007/JHEP05(2020)013
\arXiv{1911.12333} [hep-th].
%949 citations counted in INSPIRE as of 31 Mar 2026

%\cite{Penington:2019kki}
\bibitem{Penington:2019kki}
G.~Penington, S.~H.~Shenker, D.~Stanford and Z.~Yang,
``Replica wormholes and the black hole interior,''
JHEP \textbf{03} (2022), 205
%doi:10.1007/JHEP03(2022)205
\arXiv{1911.11977} [hep-th].
%1050 citations counted in INSPIRE as of 31 Mar 2026

%\cite{Hartman:2020swn}
\bibitem{Hartman:2020swn}
T.~Hartman, E.~Shaghoulian and A.~Strominger,
``Islands in Asymptotically Flat 2D Gravity,''
JHEP \textbf{07} (2020), 022
%doi:10.1007/JHEP07(2020)022
\arXiv{2004.13857} [hep-th].
%167 citations counted in INSPIRE as of 06 Mar 2026

%\cite{Leichenauer:2014nxa}
\bibitem{Leichenauer:2014nxa}
S.~Leichenauer,
``Disrupting Entanglement of Black Holes,''
Phys. Rev. D \textbf{90} (2014) no.4, 046009
doi:10.1103/PhysRevD.90.046009
\arXiv{1405.7365} [hep-th].
%103 citations counted in INSPIRE as of 08 Apr 2026

%\cite{Perlmutter:2016pkf}
\bibitem{Perlmutter:2016pkf}
E.~Perlmutter,
``Bounding the Space of Holographic CFTs with Chaos,''
JHEP \textbf{10} (2016), 069
%doi:10.1007/JHEP10(2016)069
\arXiv{1602.08272} [hep-th].
%171 citations counted in INSPIRE as of 07 Apr 2026

%\cite{Hollowood:2021wkw}
\bibitem{Hollowood:2021wkw}
T.~J.~Hollowood, S.~P.~Kumar, A.~Legramandi and N.~Talwar,
``Ephemeral islands, plunging quantum extremal surfaces and BCFT channels,''
JHEP \textbf{01} (2022), 078
%doi:10.1007/JHEP01(2022)078
\arXiv{2109.01895} [hep-th].
%24 citations counted in INSPIRE as of 31 Mar 2026

%\cite{Mertens:2022irh}
\bibitem{Mertens:2022irh}
T.~G.~Mertens and G.~J.~Turiaci,
``Solvable models of quantum black holes: a review on Jackiw{\textendash}Teitelboim gravity,''
Living Rev. Rel. \textbf{26} (2023) no.1, 4
%doi:10.1007/s41114-023-00046-1
\arXiv{2210.10846} [hep-th].
%246 citations counted in INSPIRE as of 31 Mar 2026

%\cite{Casini:2009sr}
\bibitem{Casini:2009sr}
H.~Casini and M.~Huerta,
``Entanglement entropy in free quantum field theory,''
J. Phys. A \textbf{42} (2009), 504007
%doi:10.1088/1751-8113/42/50/504007
\arXiv{0905.2562} [hep-th].
%795 citations counted in INSPIRE as of 31 Mar 2026

%\cite{Callan:1992rs}
\bibitem{Callan:1992rs}
C.~G.~Callan, Jr., S.~B.~Giddings, J.~A.~Harvey and A.~Strominger,
``Evanescent black holes,''
Phys. Rev. D \textbf{45} (1992) no.4, R1005
%doi:10.1103/PhysRevD.45.R1005
\arXiv{hep-th/9111056} [hep-th].
%1235 citations counted in INSPIRE as of 06 Mar 2026


%\cite{Calabrese:2009ez}
%\bibitem{Calabrese:2009ez}
%P.~Calabrese, J.~Cardy and E.~Tonni,
%``Entanglement entropy of two disjoint intervals in conformal field theory,''
%J. Stat. Mech. \textbf{0911} (2009), P11001
%doi:10.1088/1742-5468/2009/11/P11001
%[\arXiv{0905.2069} [hep-th]].
%355 citations counted in INSPIRE as of 20 Apr 2026

%\cite{Lowe:1992ed}
\bibitem{Lowe:1992ed}
D.~A.~Lowe,
``Semiclassical approach to black hole evaporation,''
Phys. Rev. D \textbf{47} (1993), 2446-2453
%doi:10.1103/PhysRevD.47.2446
\arXiv{hep-th/9209008} [hep-th].
%76 citations counted in INSPIRE as of 06 Mar 2026

%\cite{Piran:1993tq}
\bibitem{Piran:1993tq}
T.~Piran and A.~Strominger,
``Numerical analysis of black hole evaporation,''
Phys. Rev. D \textbf{48} (1993), 4729-4734
%doi:10.1103/PhysRevD.48.4729
\arXiv{hep-th/9304148} [hep-th].
%72 citations counted in INSPIRE as of 06 Mar 2026


%\cite{Gautason:2020tmk}
\bibitem{Gautason:2020tmk}
F.~F.~Gautason, L.~Schneiderbauer, W.~Sybesma and L.~Thorlacius,
``Page Curve for an Evaporating Black Hole,''
JHEP \textbf{05} (2020), 091
%doi:10.1007/JHEP05(2020)091
\arXiv{2004.00598} [hep-th].
%182 citations counted in INSPIRE as of 31 Mar 2026


%\cite{Callebaut:2018nlq}
\bibitem{Callebaut:2018nlq}
N.~Callebaut and H.~Verlinde,
``Entanglement Dynamics in 2D CFT with Boundary: Entropic origin of JT gravity and Schwarzian QM,''
JHEP \textbf{05} (2019), 045
%doi:10.1007/JHEP05(2019)045
\arXiv{1808.05583} [hep-th].
%27 citations counted in INSPIRE as of 31 Mar 2026


%\cite{Reyes:2021npy}
\bibitem{Reyes:2021npy}
I.~A.~Reyes,
``Moving Mirrors, Page Curves, and Bulk Entropies in AdS2,''
Phys. Rev. Lett. \textbf{127}, no.5, 051602 (2021)
%doi:10.1103/PhysRevLett.127.051602
\arXiv{2103.01230} [hep-th].
%33 citations counted in INSPIRE as of 28 Apr 2026

%\cite{Fritzsch:2021bpm}
\bibitem{Fritzsch:2021bpm}
F.~Fritzsch and T.~Prosen,
``Boundary chaos,''
Phys. Rev. E \textbf{106}, no.1, 014210 (2022)
%doi:10.1103/PhysRevE.106.014210
\arXiv{2112.05093} [cond-mat.stat-mech].
%9 citations counted in INSPIRE as of 28 Apr 2026

\bibitem{Fritzsch:2023byj}
F.~Fritzsch, R.~Ghosh and T.~Prosen,
``Boundary chaos: Exact entanglement dynamics,''
SciPost Phys. \textbf{15}, no.3, 092 (2023)
%doi:10.21468/SciPostPhys.15.3.092
\arXiv{2301.08168} [cond-mat.stat-mech].
%10 citations counted in INSPIRE as of 28 Apr 2026

%\cite{Balasubramanian:2021xcm}
\bibitem{Balasubramanian:2021xcm}
V.~Balasubramanian, B.~Craps, M.~Khramtsov and E.~Shaghoulian,
``Submerging islands through thermalization,''
JHEP \textbf{10}, 048 (2021)
%doi:10.1007/JHEP10(2021)048
\arXiv{2107.14746} [hep-th].
%33 citations counted in INSPIRE as of 28 Apr 2026

%\cite{Liu:2020gnp}
\bibitem{Liu:2020gnp}
H.~Liu and S.~Vardhan,
``A dynamical mechanism for the Page curve from quantum chaos,''
JHEP \textbf{03} (2021), 088
%doi:10.1007/JHEP03(2021)088
\arXiv{2002.05734} [hep-th].
%61 citations counted in INSPIRE as of 29 Apr 2026

%\cite{Gu:2017njx}
\bibitem{Gu:2017njx}
Y.~Gu, A.~Lucas and X.~L.~Qi,
``Spread of entanglement in a Sachdev-Ye-Kitaev chain,''
JHEP \textbf{09} (2017), 120
%doi:10.1007/JHEP09(2017)120
\arXiv{1708.00871} [hep-th].
%104 citations counted in INSPIRE as of 29 Apr 2026

%\cite{Chen:2017dbb}
\bibitem{Chen:2017dbb}
Y.~Chen, H.~Zhai and P.~Zhang,
``Tunable Quantum Chaos in the Sachdev-Ye-Kitaev Model Coupled to a Thermal Bath,''
JHEP \textbf{07} (2017), 150
%doi:10.1007/JHEP07(2017)150
\arXiv{1705.09818} [hep-th].
%68 citations counted in INSPIRE as of 29 Apr 2026

%\cite{Chen:2020wiq}
\bibitem{Chen:2020wiq}
Y.~Chen, X.~L.~Qi and P.~Zhang,
``Replica wormhole and information retrieval in the SYK model coupled to Majorana chains,''
JHEP \textbf{06} (2020), 121
%doi:10.1007/JHEP06(2020)121
\arXiv{2003.13147} [hep-th].
%85 citations counted in INSPIRE as of 29 Apr 2026
 
\end{thebibliography}
\end{document}